\documentclass[12pt]{article}
\usepackage[applemac]{inputenc}
\usepackage{amsmath,amssymb}
\usepackage{paralist}
\usepackage{slashed}
\usepackage{graphicx}
\usepackage{subfigure}
\usepackage{color}
\usepackage{fullpage}
\usepackage{multirow}
\usepackage{hyperref}
\usepackage{amscd}
\usepackage{xcolor}
\usepackage{braket}
\usepackage{amsfonts}

\catcode`\@=11 
\def\numberbysection{\@addtoreset{equation}{section} 
        \def\theequation{\thesection.\arabic{equation}}} 
\def\be{\begin{equation}} 
\def\ee{\end{equation}} 
\def\ba{\begin{eqnarray}} 
\def\ea{\end{eqnarray}} 
\def\bali{\begin{align}}
\def\eali{\end{align}}
 
\def\ov{\overline}

\def\Z{\mathbb{Z}}

\def\nl{\nonumber \\}

\def\de{\partial} 
\def\wt{\widetilde}

\def\dag{\dagger} 
 
\def\a{\alpha} 
\def\b{\beta} 
\def\g{\gamma} 
\def\G{\Gamma} 
\def\D{\Delta} 
\def\d{\delta} 
\def\e{\epsilon} 
\def\eps{\varepsilon} 
\def\z{\zeta}

\def\l{\lambda} 
\def\L{\Lambda} 
\def\m{\mu} 
\def\n{\nu}

\def\p{\pi} 
\def\P{\Pi} 
\def\r{\rho} 
\def\s{\sigma} 
\def\S{\Sigma} 
\def\t{\tau} 
\def\f{\phi} 
\def\vf{\varphi} 
\def\F{\Phi} 
\def\c{\chi} 
\def\w{\omega} 
\def\W{\Omega} 
\def\Th{\Theta} 
\def\th{\theta}

\def\En{E_{\vec{n}}}
\def\kn{\vec{k}_{\vec{n}}}

\begin{document} 
 
\begin{titlepage} 
\begin{center} 

\vskip .6 in 
{\LARGE Three-dimensional Topological Insulators}
\medskip

{\LARGE and Bosonization} 
 
\vskip 0.2in 
Andrea CAPPELLI${}^{(a)}$,\ \ \ 
Enrico RANDELLINI${}^{(a,b)}$, \ \ \ 
Jacopo SISTI${}^{(c)}$ 
 \bigskip

{\em ${}^{(a)}$ INFN, Sezione di Firenze}\\ 
{\em ${}^{(b)}$ Dipartimento di Fisica e Astronomia, Universit\`a di Firenze}\\ 
{\em Via G. Sansone 1, 50019 Sesto Fiorentino - Firenze, Italy} \\
{\em ${}^{(c)}$ SISSA, Via Bonomea 265, 34136 Trieste, Italy}
\end{center} 
\vskip .2 in 

\begin{abstract} 
Massless excitations at the surface of three-dimensional time-reversal
invariant topological insulators possess both fermionic and bosonic
descriptions, originating from band theory and hydrodynamic BF
gauge theory, respectively.  We analyze the corresponding
field theories of the Dirac fermion and compactified boson and compute
their partition functions on the three-dimensional torus geometry. 
We then find some non-dynamic exact properties of bosonization in
$(2+1)$ dimensions, regarding fermion parity and spin sectors.  Using
these results, we extend the Fu-Kane-Mele stability argument to
fractional topological insulators in three dimensions.
\end{abstract} 
 
\vfill 

\end{titlepage} 
\pagenumbering{arabic}
\numberbysection


\section{Introduction}

The topological phases of matter 
\cite{Fradkin-book,Kane-rev,Zhang-rev,Bernevig-book}
have been described by several
approaches, such as wavefunction modeling \cite{Hansson-wf}, 
band theory \cite{Class} and effective
field theory of boundary excitations \cite{Wen-book,Moore-BF},
whose interplay has been extremely rich and fruitful.
In this paper, we analyze $(3+1)$-dimensional time-reversal
invariant topological insulators using field theory methods.

The main motivation of our study is the success of the field theory approach
for $(2+1)$ dimensional topological states, where
exact methods are available for describing the one-dimensional
edge excitations, most notably those of conformal field theory \cite{cft}.
In several instances, these methods give access to strongly interacting
dynamics and make use of powerful symmetry principles.
The extensive modeling of quantum Hall states has been applied
to the description of the quantum spin Hall effect and then to time-reversal
invariant topological insulators \cite{LS}. 

In particular, the $\Z_2$ characterization of stability of topological
insulators, originally derived within band theory by Fu, Kane and Mele
\cite{Kane-Z2, Moore-Z2}, has been reformulated in field-theory language and
extended to interacting fermion models with Abelian \cite{LS} and
non-Abelian \cite{cr1,cr2} fractional statics of excitations.
The $\Z_2$ stability also extends to $(3+1)$ dimensional band
insulators and it is interesting to find the corresponding field
theory argument for analyzing interacting systems.  In this paper, we
shall present results in this direction.

More generally, the theoretical methods in $(3+1)$ dimensions
are facing the problem of bosonization, namely that of finding
correspondences between two seemingly different theoretical approaches:
\begin{itemize}
\item
That of fermionic theories, dealing with band structures and topological
effects related to Berry phases, and leading to
the ten-fold classification of non-interacting topological states
\cite{Class}.
\item
That of bosonic theories, also called hydrodynamic approach, dealing
with topological gauge theories and their description of braiding
relations and boundary excitations 
\cite{Hansson-BF,TI-qft,Moore-BF,Magnoli-BF,Fradkin-BF}.
\end{itemize}

Bosonization is an exact map in $(1+1)$-dimensional field
theories that is very well understood \cite{cft}; thus, 
the above interplay does not cause any problem for $(2+1)$-dimensional
topological states. The bosonic approach can provide exact
results for interacting systems and well as
the methods for discussing bulk wavefunctions and braiding statistics.

In this paper, we review and develop both the fermionic and bosonic
field theory descriptions of massless surface states for time-reversal 
invariant topological insulators in $(3+1)$ dimensions. 
Our main method is the study of partition functions on the space-time
geometry of the three-torus and their behaviour
under flux insertions and modular transformations, namely for
large gauge transformations of the electromagnetic and gravitational
backgrounds \cite{cz,cr1}.
In the fermionic theory, we study the free Dirac excitations at the
surface of topological insulators \cite{Ryu-F}. In the bosonic approach, we
analyze the BF topological theory and the associated bosonic field theory
of surface excitations \cite{Moore-BF,Ryu-B}. We then quantize the theory 
of the compactified boson.

Although these fermionic and bosonic theories are different
in $(2+1)$ dimensions, we can establish some basic properties
of bosonization that are exact being independent of dynamics:
\begin{itemize}
\item 
We show that the quantization of the compactified boson in $(2+1)$
dimension yields eight sectors that correspond to the spin sectors of
the fermionic theory on the three-torus. The partition
functions in the two theories transform in the same way under flux insertions
and modular transformations; actually, they become equal under dimensional
reduction to $(1+1)$ dimensions, where bosonization is an exact map.
\item
We identify the fermion parity of bosonic states and using
this assignment we formulate the Fu-Kane-Mele stability argument
in the bosonic theory. 
\item
We then prove the stability of interacting topological insulators
described by the hydrodynamic BF theory with odd integer coupling
constant $K$, possessing fractional changed quasiparticles and
fractional Abelian braiding between quasiparticles and
vortices.
\end{itemize}

The paper includes the following parts. In Section 2, we recall the
fermionic theory of surface states of topological
insulators, compute the partition function on the three-torus and
study its transformations properties. We reformulate the
Fu-Kane-Mele stability analysis in terms of properties of the spectrum
in the different spin sectors and study the dimensional reduction to
the two-torus. In section 3, we review the hydrodynamic approach of the
BF theory, derive the bosonic surface theory and discuss two
Hamiltonians consistent with the topological data. We then quantize the
compactified boson and obtain the partition functions.  In
Section 4, we find their transformation properties, match the
fermionic and bosonic sectors, assign the fermion parity 
and extend the stability analysis to the bosonic theory.


\section{Fermionic topological insulators}


\subsection{Introduction: surface states and anomaly cancellation}

We start by recalling some known features of three-dimensional
topological band insulators, involving non-interacting fermions with
time reversal invariance \cite{Kane-rev}.  At the
microscopic level, the topological states occur for an odd number of
level crossings between the valence and conduction bands.  Near each
crossing a Dirac cone is approximatively present and a massless
relativistic fermionic excitation is realized at small
energies. Actually, this is a $(2+1)$-dimensional Dirac fermion
located at the surface of the system \cite{Class, TI-micro}.

\begin{figure}[t]
\begin{center}
\includegraphics[width=9cm]{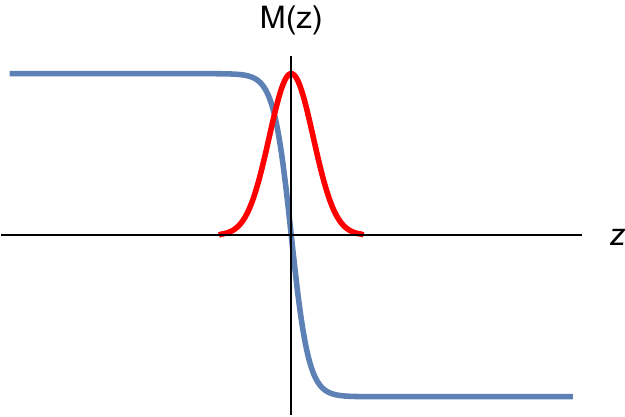}
\caption{Mass profile $M=M(z)$ (blue line) and wavefunction
of the Hamiltonian zero mode (red line).}
\label{mass-kink}
\end{center}
\end{figure}

The low-energy effective field theory description of these topological
states is realized by the $(3+1)$-dimensional 
free Dirac fermion with mass $M$ in the
bulk that vanishes at the surface \cite{Class,TI-qft}.  
Let us consider a plane located at
$z=0$ separating the bulk of the material ($z<0$) from
empty space ($z>0$): we can take the mass profile 
$M(z)=-M_0\tanh(z/\ell)$, where $M_0$ is of the order of the bulk gap
and $\ell$ of the lattice spacing (see Fig.\,\ref{mass-kink}).

The Dirac theory with this mass profile possesses massless fermionic
surface excitations that are obtained by the so-called Jackiw-Rebbi
dimensional reduction, originally formulated for the polyacetylene chain
in $(1+1)$ dimensions \cite{JR, Fradkin-book}. 
Let us recall the main steps of this
argument in the case of reduction from $(3+1)$ to $(2+1)$ dimensions. 
A convenient basis for the $\g$ matrices in $d=(3+1)$ is given by:
\begin{equation}
\label{rep2}
\gamma^0= \left(\begin{matrix}
0 & \sigma_3 \\
\sigma_3 & 0
\end{matrix}\right), \ \ 
\gamma^1= i\left(\begin{matrix}
0 & \sigma_1 \\
\sigma_1 & 0
\end{matrix}\right),  \ \ 
\gamma^2= i\left(\begin{matrix}
0 & \sigma_2 \\
\sigma_2
 & 0 \end{matrix}\right),  \ \ 
\gamma^3= i\left(\begin{matrix}
1 & 0 \\
0 & -1
\end{matrix}\right),
\end{equation}
where  the $\s$'s are the Pauli matrices. 
The Dirac Hamiltonian takes the form\footnote{
Throughout this paper we shall use natural units $c=\hbar=1$,
normalize the Fermi velocity of massless excitations to one and
set the electric charge $e=1$.
}:
\be
\label{dirac-red}
H=-i \gamma^0 \gamma^1 \partial_{x}-i\gamma^0 \gamma^2 \partial_{y}-
i \gamma^0 \gamma^3 \partial_{z} +\gamma^0 M(z) \equiv H_{xy}+H_z.
\ee
We look for eigenstates of  $H$ which are simultaneously zero-energy 
eigenstates of $H_z$, so as to realize the dimensional reduction.
We assume the separation of variables,
\be
\label{sep-var}
H\psi=E\psi, \qquad\quad
\psi=\vf(z)u(x,y),
\ee
where $\vf(z)$ is a function and $u(x,y)$ a spinor, and impose the
zero-energy condition for $H_z$, i.e.
\begin{equation}
\label{e2}
\left( i \partial_z + \gamma^3 M(z) \right)\vf(z)u(x,y) =0.
\end{equation}
The solutions to this equation are of the form:
\begin{equation}
\label{e1} 
\psi_{\pm}=\vf_\pm(z)u_\pm, \qquad\quad
\left( \partial_z \pm M(z) \right)\vf_\pm(z)=0\ ,
\qquad\quad \gamma^3 u_\pm=\pm i u_\pm\ .
\end{equation}
In this equation, the spinor $u_\pm$ has non-vanishing components in
the upper/lower two-dimensional subspaces of the representation 
(\ref{rep2}).
Only one solution of (\ref{e1}) is normalizable, 
given the kink shape in Fig.\,\ref{mass-kink}, namely $\vf_-$:
\begin{equation}
\label{fi-}
\vf_-(z)=\exp\left(\int_0^z dz'~ M(z')\right).
\end{equation}
The explicit calculation shows that this zero mode is localized at the 
surface $z=0$ where the mass changes sign.

The remaining surface dynamics is governed by the Hamiltonian 
$H_{xy}$ acting on spinors of the form $u_-=(0,\c_-)$, 
where $\c_-$ is a bicomponent spinor. Once projected 
in this subspace, $H_{xy}$ takes the form,
\begin{equation}
(k_y\sigma_1-k_x\sigma_2)\chi_-=E\chi_-\ ,
\label{dirac_2+1}
\end{equation}
in terms of surface momenta $(k_x,k_y)$.
This is the expected Hamiltonian of a massless Dirac excitation in
$(2+1)$ dimensions. Of course, the dimensional reduction 
holds for energies $E\ll M_0$.

In a physical setup, the system has two boundaries along the $z$
axis, with the second surface described by the inverted mass
profile $M(z) \to -M(z+z_0)$. Performing the same steps as before, the
normalizable zero mode is now given by the positive sign in 
(\ref{e1}), i.e. $u_+=(\chi_+, 0)$. It turns out that the 
bispinor $\c_+$ obeys the same Dirac equation \eqref{dirac_2+1}.

In the study of the microscopic model of band topological insulators,
the authors of Ref.\cite{TI-micro} considered the expansion of
the Hamiltonian for momenta near the band crossing point and obtained
the result (\ref{dirac_2+1}) at low-energy. In that approach, the
angular momentum of the surface states was also discussed, involving
the electron spin as well as a $L=1$ contribution from the $p$-wave
orbitals involved.  They found that the low-energy surface excitations
have angular momentum one-half that is represented by $S_i\sim \s_i$, where the
$\s_i$ are the Pauli matrices introduced before.  Therefore, the
solutions of (\ref{dirac_2+1}) have associated the values:
\be
\label{heli}
\langle S_z \rangle=0\ ,
\qquad\quad
\langle S_x k_x +S_y k_y \rangle=0\ .
\ee
Namely, the spin lies on the surface and is orthogonal to momentum
(helical spin excitations).
These results are in agreement with the time reversal symmetry
${\cal T}$ in $(3+1)$ dimensions: the Dirac mass is ${\cal T}$
invariant, as well as the helical states on the boundary of the
topological insulator.

We remark that the analogous Jackiw-Rebbi dimensional reduction from
$(2+1)$ to $(1+1)$ dimensions would lead to chiral (resp. antichiral)
edge fermions for kink (resp. antikink) mass profile. Indeed, the
massive Dirac theory in $(2+1)$ dimensions breaks parity and time reversal
symmetries and can model chiral topological states such as the (anomalous)
quantum Hall effect and Chern insulators \cite{Fradkin-book}.  
Analogous Jackiw-Rebbi
reductions exist for all ten classes of non-interacting fermionic
topological states, and actually provide an independent
derivation of the classification \cite{Class}.

Fermionic surface states with mass $\pm m$ can be introduced by adding
the following ${\cal T}$ breaking mass to the $(3+1)$ dimensional 
Dirac Hamiltonian (\ref{dirac-red}):
\be
\label{5-mass}
\Delta H= i\g^0\g^5 m\ .
\ee
Upon repeating the Jackiw-Rebbi reduction, one find the following
$(2+1)$-dimensional Dirac Lagrangians on the two boundary surfaces,
\be
\label{lagrJR}
\mathcal{L}_-=\ov{\c}_- (i \slashed \de  - m )\c_-, 
\qquad\quad
\mathcal{L}_+=\ov{\c}_+ (i \slashed \de  + m )\c_+.
\ee
Time reversal transformations act as ${\cal T}: m \to -m$;
moreover, the two electrons correspond to inequivalent representations
of the Clifford algebra, such that the mass sign is physically
relevant for surface fermions.

Another result that is relevant for the following discussion
is the induced effective action obtained by coupling  the $(2+1)$-dimensional 
fermion to the electromagnetic field $A_\mu$.
The expansion of the fermionic determinant 
$\det\left(i\slashed \de+ \slashed A -m \right)$ corresponds to
an effective action involving any power of $A_\mu$. The first, 
quadratic term is given by the one-loop vacuum polarization
$\P_{\m\n}$,
\be
\label{2indA}
S_{\rm eff}[A]=-\frac{1}{2} \int \frac{d^3k}{(2\p)^3} 
\ A_{\m}(k)\P_{\m\n}(k,m)A_{\n}(-k)\ + \ O\left(A^3 \right)\ .
\ee
This reads in Euclidean space \cite{JT, Redlich}:
\begin{align}
     \P_{\m\n}(k,m)=& \frac{1}{4\p} k_\a \e^{\a\m\n}\left(
     \frac{m}{|m|} \frac{\arctan(x)}{x} -\frac{\L}{|\L|} \right) 
\notag
\\ &
     - \left(k^2 \d_{\m\n} -k_\m k_\n\right) \frac{1}{8\pi |k|}
     \left(\frac{1}{x}- \frac{1-x^2}{x^2}\arctan(x) \right), 
\notag\\
    x=&\frac{|k|}{2|m|}\ .
\label{Ploop1}
\end{align} 
This expression contains a first term that is odd in
momentum, and a second that is even.  The first term breaks parity and time
reversal, and has been regularized by subtracting the contribution of
a Pauli-Villars fermion with mass $\L\to\infty$.

In the limit of large mass $|m|\to\infty$, the expression for
$\Pi_{\mu\nu}$ becomes:
\be
\label{Ploop2}
     \P_{\m\n}(k,m)=
\frac{1}{4\p} k_\a \e^{\a\m\n}\left(
\frac{m}{|m|}-\frac{\L}{|\L|} \right) 
  - \left(k^2 \d_{\m\n} -k_\m k_\n\right) \frac{1}{12\pi |m|} \ ,
\quad (|m|\gg |k|).
\ee
The vanishing of the effective action in the static limit
fixes the sign of the regulator mass to be ${\rm sign}(\L)={\rm sign}(m)$.
In the massless limit, one instead obtains the result
\be
\label{Ploop3}
     \P_{\m\n}(k,m)=
-\frac{1}{4\p} k_\a \e^{\a\m\n} \frac{m}{|m|} 
  - \left(k^2 \d_{\m\n} -k_\m k_\n\right) \frac{1}{16 |k|},
\qquad (|m|\ll |k|).
\ee
The first term corresponds to an induced Chern-Simons action,
\be
\label{ind-CS}
S_{CS}[A] = i\frac{K}{4\pi}
\int d^3x\, \eps^{\mu\nu\r}
A_\mu\de_\nu A_\r, \qquad\quad
K=\frac{1}{2}\frac{m}{|m|}\ .
\ee

Therefore, in the massless theory the parity and time reversal
symmetries are restored but are broken at the quantum level: this is
the $\Z_2$ anomaly of $(2+1)$-dimensional fermions 
\cite{Redlich}.
Note that the sign of Chern-Simons coupling $K$, 
i.e. the sign of the Pauli-Villars regulator, cannot be determined
in the massless theory without referring to a massive phase \cite{Seminara}. 
The second, parity invariant term in the effective action (\ref{Ploop1})
will be relevant for the discussion in Section 3.2.

In the case of topological insulators, the anomaly (\ref{ind-CS}) of surface
massless states cancels against the bulk contribution.
Let us recall how this occurs.
The bulk action is given by the Abelian theta term \cite{TI-qft},
\be
S_\theta[A]= -\frac{\theta}{32 \pi^2} \int d^4x ~ 
\epsilon^{\mu\nu\lambda \rho} F_{\mu\nu}F_{\lambda \rho},
\label{theta}
\ee
with parameter $\th=\pi$.
In the geometry of the slab considered here, $S_\theta[A]$ is a total derivative
that reduces to the Chern-Simons action (\ref{ind-CS}) on the two surfaces
with coupling taking opposite values $K=\pm 1/2$.

Furthermore, in the earlier discussion of the Jackiw-Rebbi reduction
we argued that the $(3+1)$-dimensional time-reversal breaking mass
induces opposite masses on the two surfaces, Eqs. (\ref{5-mass}),
(\ref{lagrJR}).  It is thus natural to expect that in the $m\to 0$
limit, the corresponding anomalies are given by Chern-Simons actions
$S_{CS}^+[A]$ and $S_{CS}^-[A]$ with opposite couplings $K_-=1/2$ and
$K_+=-1/2$.
Therefore, the time-reversal breaking terms cancel in the whole 
system by matching their signs, as follows:
\be
S_{\rm tot}[A]=S^+_{CS}[A]+ S_\theta[A]+S^-_{CS}[A] =0\ ,
\qquad\quad \theta=\pi.
\label{bulk-bordo}
\ee

This result establishes the bulk-boundary cancellation of the $\Z_2$
anomaly in $(3+1)$-dimensional topological insulators \cite{Burnell,
Witten}.  Actually, it should rather be called a
boundary-boundary cancellation: the theta term does not imply local
effects in the bulk; its role is that of `transporting' the anomalous
action $S_{CS}[A]$ from one surface to the other.  This result should
be contrasted with the mechanism of `anomaly inflow' \cite{Inflow} in the
quantum Hall effect in $(2+1)$ dimensions, where the chiral anomaly 
of the edge, i.e. the non-conservation of charge, is compensated 
by the classical Hall current in the bulk \cite{Fradkin-book,cdtz}.
 
In the case of compact $(3+1)$-dimensional 
manifolds, $S_{\rm tot}[A]=S_\th[A]$ does not
vanish, but is proportional to the integral of the second Chern class
${\cal C}_2$ of the gauge field, $S_\th[A]=\th\, {\cal C}_2$. Since
${\cal C}_2$ is an integer valued topological invariant quantity, the coupling
$\th$ is defined modulo $2\pi$.  Therefore, $S_{\rm tot}[A]$ is
time reversal invariant due to the equivalence of the values $\th=\pi$
and $\th=-\pi$ \cite{TI-qft}.

Altogether, topological insulators are time reversal invariant
topological states of matter. Their stability with respect to
interactions and perturbations is not guaranteed and should be checked
carefully.  If time reversal symmetry is broken they decay into 
trivial insulators, i.e. they belong to the class of symmetry protected
topological states, to be contrasted with other states, such as Chern
insulators, that are chiral and stable \cite{Zhang-rev, Wen-spt}.
Topological insulators are characterized by a $\Z_2$ index of
stability, that is zero for unstable/trivial
insulators and one for stable/topological insulators 
\cite{Kane-Z2, Moore-Z2}.  At the level of
band theory, this index distinguishes the cases of even number of
level crossings, that can be smoothly deformed to no crossing, from
that of odd crossings, that can be deformed to one crossing. Within
the effective theory, we should consider even and odd
numbers of Dirac surface fermions  \cite{Zhang-rev, TI-qft}. 
Since a time-reversal invariant quadratic (mass)
interaction can be written in terms of two fermion species, 
it can be used to gap them in pairs, thus remaining with zero or 
one massless fermion.  

In $(2+1)$-dimensional topological insulators, the $\Z_2$ index of
stability also holds for systems of interacting fermions, including
Abelian \cite{LS} as well as non-Abelian \cite{cr1,
  cr2} fractional insulators.  Stability in $(3+1)$-dimensional
interacting states is not yet fully understood and some results will
be presented in the following sections of the paper.

We finally discuss the explicit breaking of time reversal symmetry at
the edge, for example due to proximity with a magnetic material \cite{TI-qft}.
In this case, the surface fermion acquires a mass and
its effective action is given by the expression 
(\ref{Ploop1}) with ${\rm sign}(\L)={\rm sign}(m)$.
In the low energy limit, $|k|\ll |m|$, 
 the anomalous term vanishes, as shown by (\ref{Ploop2}). It follows that
the bulk theta term $S_\th[A]$ is not cancelled as in (\ref{bulk-bordo})
and it reduces to (minus) the Chern-Simons action (\ref{ind-CS}) at the surface.
This implies a surface Hall effect with conductivity
$\sigma_H=e^2\nu/h$, $\nu= 1/2$ \cite{Hasan-exp}.


\subsection{Torus partition functions}

In this section, we derive the partition function of the Dirac
fermion on the surface of the $(3+1)$-dimensional topological 
insulator. Extending our earlier analysis in
one less dimension \cite{cr1}, we use the partition function
to reformulate the Fu-Kane-Mele stability argument for the existence of
massless surface states in presence of disorder and
interactions that are time reversal invariant (the `strong
topological insulators' of Ref.\cite{Kane-Z2}).
These results are also the starting point for
discussing the stability of bosonic surface states in Section 4.

\begin{figure}[t]
\begin{center}
\includegraphics[width=10cm]{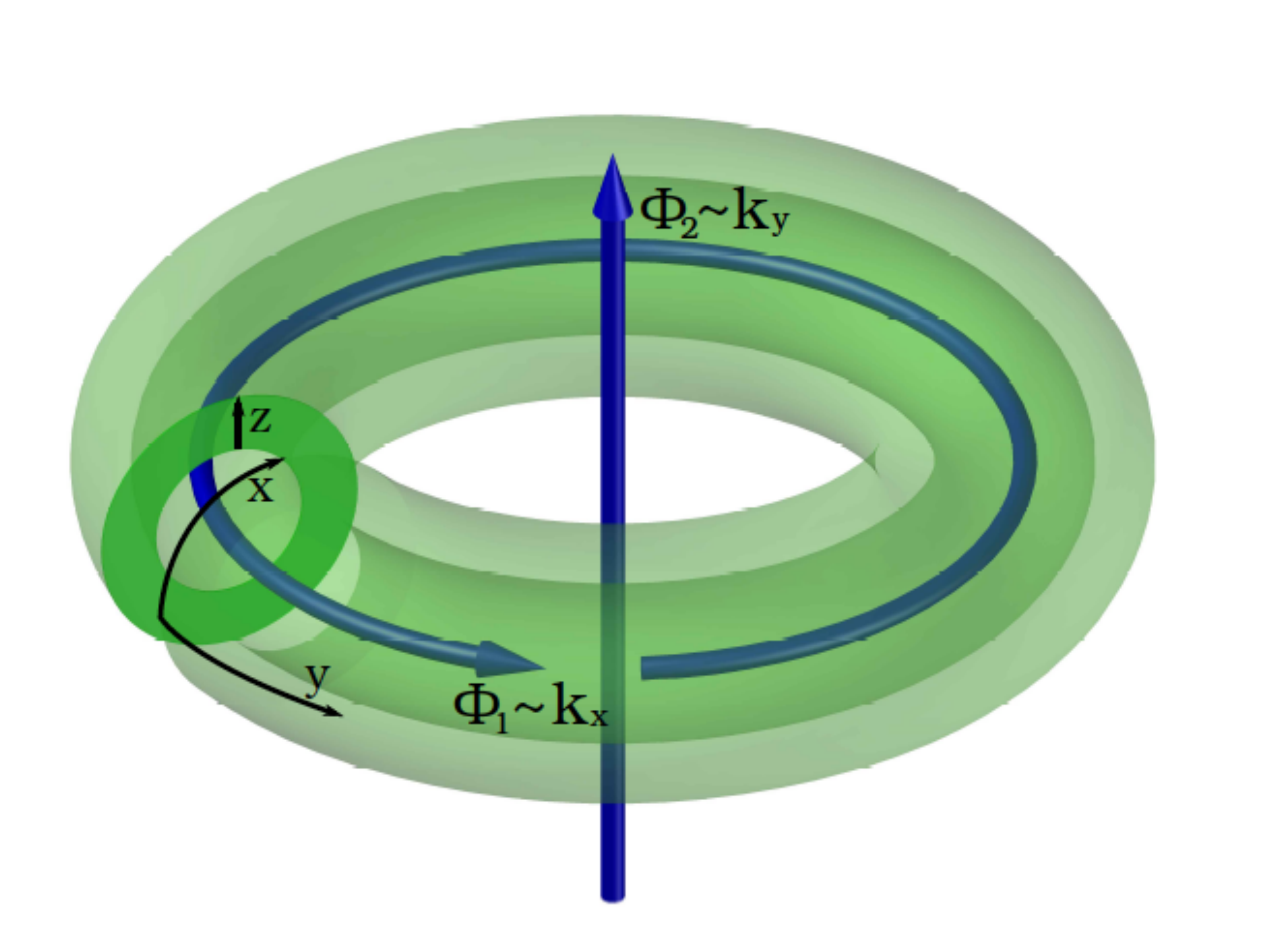}
\caption{Three-dimensional Corbino donut. The addition of fluxes
  $\F_1$ and $\F_2$ modifies the quantization of $k_x$ and $k_y$
  momenta, respectively.}
\label{T3flussi}
\end{center}
\end{figure}

We consider the spatial geometry of a `Corbino donut' (see
Fig.\,\ref{T3flussi}), whose internal and external surfaces are
two-torii. The space-time three-torus $\mathbb{T}^3$ is obtained by
considering one boundary and the Euclidean time with
period the temperature $T$.  The partition functions for periodic $(P)$ and
anti-periodic $(A)$ boundary conditions in space and time form a
eight-dimensional multiplet, corresponding to the eight spin sectors
of $\mathbb{T}^3$. Our derivation of the partition functions uses results of
Ref.\cite{cc}\cite{Ryu-F}.

\begin{figure}[t]
\begin{center}
\includegraphics[width=15cm]{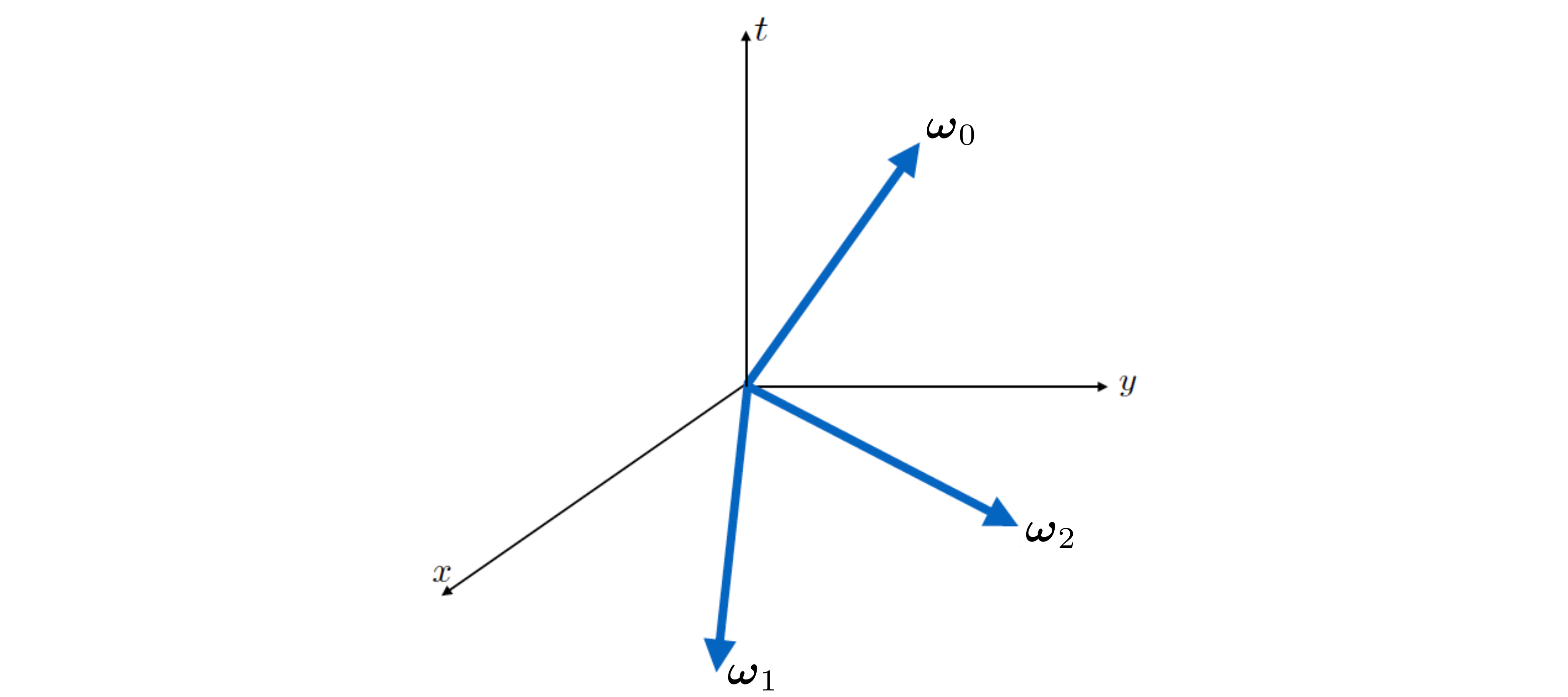}
\caption{Vector moduli of $\mathbb{T}^3$ torus in 
$(2+1)$ dimensions.}
\label{torus}
\end{center}
\end{figure}

The $\mathbb{T}^3$ torus is defined by the 
 generators of the periodicity lattice $\boldsymbol{\w}_\m, \m=0,1,2$,
(see Fig.\,\ref{torus}), whose components form
the following matrix:
\be
\label{omegagen}
\begin{pmatrix}
\boldsymbol{\w}_0 \\
\boldsymbol{\w}_1\\
\boldsymbol{\w}_2
\end{pmatrix}=
\begin{pmatrix}
\w_{00} & \w_{01} & \w_{02} \\
0  & \w_{11} & \w_{12} \\
0  & \w_{21}  & \w_{22}
\end{pmatrix}\ .
\ee
The dual vectors $\boldsymbol{k}_{\m}$ are defined by 
$ \boldsymbol{k}_{\m}\cdot \boldsymbol{\w}_{\n}=\d_{\m\n}$; 
their spatial components will be indicated as:
\be
\label{2d-moduli}
\vec{\w}_i=(\w_{1i},\w_{2i}), \ \ \ \ \ \ \vec{k}_i=(k_{1i},k_{2i}),
\qquad i=1,2.
\ee
The volumes $V^{(3)}$ and $V^{(2)}$ of the $3D$ space-time and  
$2D$ space cells are respectively given by:
\be
\label{vol}
V^{(3)}=\text{det}(\boldsymbol{\w}), \ \ \ \ 
V^{(2)}=\text{det} \ \vec{\w}=|\boldsymbol{\w}_1\times \boldsymbol{\w}_2|.
\ee
Some useful relations are:
\be
\label{k-to-w}
k_{11}=\frac{\w_{22}}{V^{(2)}},\ \   \ k_{12}=-\frac{\w_{21}}{V^{(2)}}, 
\ \ \  k_{21}=-\frac{\w_{12}}{V^{(2)}}, \ \ \ k_{22}=\frac{\w_{11}}{V^{(2)}},
\quad \w_{00}=\frac{V^{(3)}}{|\boldsymbol{\w}_1\times \boldsymbol{\w}_2|}\ .
\ee

The partition function is given in terms of the on-shell data of
the free fermion, namely its spectra of energy, momentum, charge
and fermion number. The usual creation and annihilation operators 
of particles $(a_{\vec{ n}}^{\dag}, a_{\vec{ n}})$ 
and antiparticles $(b_{\vec{ n}}^{\dag}, b_{\vec{ n}})$ ,
where $\vec{ n}=(n_1,n_2) \in\mathbb{Z}^2$, obey
anti-commutation relations and satisfy the vacuum conditions
\be
\label{vac-cond}
a_{\vec{ n}} \ket{\Omega}=b_{\vec{ n}}\ket{\Omega}=0 \ ,
\qquad \quad n_1,n_2 \in \Z\ .
\ee
The energy, momentum, charge and fermion number of excitations 
are given by the following normal ordered expressions:
\begin{align}
\label{spec-e}
    & E=\sum_{\vec{ n}}  E_{\vec{n}} \left( a^{\dag}_{\vec{ n}}a_{\vec{ n}}+
b_{\vec{ n}}^{\dag}b_{\vec{ n}} -1 \right)\ , \qquad
 E_{\vec{n}}=2\pi\big| (n_1+\a_1)\vec{k}_1 + (n_2+\a_2)\vec{k}_2 \big|,  
\\
\label{spec-p}    
&  P_i=\sum_{\vec{ n}} p_{\vec{n}} \left( a^{\dag}_{\vec{ n}}a_{\vec{ n}}+
b_{\vec{ n}}^{\dag}b_{\vec{ n}} \right), \qquad
\quad p_{\vec{n}}=2\pi\left( (n_1+\a_1)k_{1i} + (n_2+\a_2)k_{2i} \right),  
\\
\label{spec-q}
& Q= \sum_{\vec{ n}}  a^{\dag}_{\vec{ n}}a_{\vec{ n}}-
b_{\vec{ n}}^{\dag}b_{\vec{ n}},
\\
\label{spec-f}& 
(-1)^F=(-1)^{\sum_{\vec{ n}}  a^{\dag}_{\vec{ n}}a_{\vec{ n}}+
b_{\vec{ n}}^{\dag}b_{\vec{ n}}}.
\end{align}
In the expression of the energy, the infinite sum $(- \sum_{\vec{n}}
E_{\vec{n}})$ to be regularized later yields the Casimir energy and
the parameters $\a_i$ specify the boundary conditions along the
$i=1,2$ spatial directions: $\a_i=0$ (resp. $\a_i=1/2$) for periodic
(P) conditions (resp.  anti-periodic (A)) conditions. For convenience,
we shall also denote the P (resp. A) conditions in time by using
another parameter $\a_0=0$ (resp. $\a_0=1/2$).

The partition function is evaluated in presence of a scalar
potential $A_0$ coupling to charge. For $(A)$ boundary conditions
in time, it is given by the following expression:
\be
\label{Z1}
Z^F_{\frac{1}{2},\a_1\a_2}=\text{Tr}\bigg[ \text{exp}\bigg( -T(E+QA_0) + 
i\w_{01} P_1+i\w_{02} P_2 \bigg) \bigg].
\ee
For  $(P)$ boundary conditions in time, the trace includes 
the  operator $(-1)^F$:
\be
\label{Z2}
Z^F_{0,\a_1\a_2}=\text{Tr}\bigg[ (-1)^F\text{exp}\bigg( -T(E+QA_0) +
  i\w_{01} P_1+i\w_{02} P_2 \bigg) \bigg].  
\ee 
The trace over the fermionic Fock space is straightforward.
In the result, we substitute the $\vec{k}_i$ vectors in terms of the
$\vec{\omega}_i$ and introduce the constant $\a_0=0,1/2$, to obtain:
\ba
\label{Z-alphai}
Z^F_{\alpha_0,\alpha_1 \alpha_2} 
&=&  e^{-V^{(3)}F_0} \prod_{n_1,n_2\in \mathbb{Z}} 
\left\{ 1-\text{exp}\left( -2 \pi  \mathcal{E}^{\a_1 \a_2}_{n_1 n_2} +
2\pi i \mathcal{P}^{\a_1 \a_2}_{n_1 n_2} -2 \pi i \mathcal{A} \right)\right\}
\nl 
& &\qquad\qquad\ \ \times
\left\{ 1-\text{exp}\left( -2 \pi  \mathcal{E}^{\a_1 \a_2}_{n_1 n_2} 
-2\pi i \mathcal{P}^{\a_1 \a_2}_{n_1 n_2} +2 \pi i \mathcal{A} \right)\right\}.
\label{partiz-funz-3d}
\ea
In this expression,
\begin{eqnarray}
\label{Z-A}
 \mathcal{A}&=&\a_0 -i \frac{V^{(3)} A_0}{2\p |{\boldsymbol{\omega}}_1 
\times{\boldsymbol{\omega}}_2|},
\\
\label{Z-E}
\mathcal{E}^{\a_1 \a_2}_{n_1 n_2}&=&\frac{ V^{(3)}}{|{\boldsymbol{\omega}}_1 
\times {\boldsymbol{\omega}}_2|^2} |(n_1+\alpha_1){\boldsymbol{\omega}}_2 - 
(n_2+\alpha_2){\boldsymbol{\omega}}_1|, \label{energie} 
\\
\label{Z-P}
\mathcal{P}^{\a_1 \a_2}_{n_1 n_2}&=& \frac{({\boldsymbol{\omega}}_1 
\times {\boldsymbol{\omega}}_2)}{|{\boldsymbol{\omega}}_1 
\times{\boldsymbol{\omega}}_2|^2} \left[ (n_1+\alpha_1)({\boldsymbol{\omega}}_0
\times {\boldsymbol{\omega}}_2)-(n_2+\alpha_2)({\boldsymbol{\omega}}_0 
\times {\boldsymbol{\omega}}_1) \right], 
\\
\label{Z-F}
 F_0&=&\frac{1}{2\pi} \sum_{n_1,n_2}{}'~ 
\frac{e^{-2\pi i (\alpha_2 n_1-\alpha_1 n_2)}}
{|n_1 {\boldsymbol{\omega}}_2-n_2 {\boldsymbol{\omega}}_1|^3}, 
\label{termine-casimir}
\end{eqnarray}
where $F_0$ is the Casimir energy regularized  
by using Epstein's analytic continuation formula (see Refs.\cite{cc}),
with the sum $\sum{}'$ excluding the value $(n_1,n_2)=(0,0)$.


\subsection{Flux insertions and stability argument}
\label{stability3d}

\begin{figure}[h]
\begin{center}
\includegraphics[width=10cm]{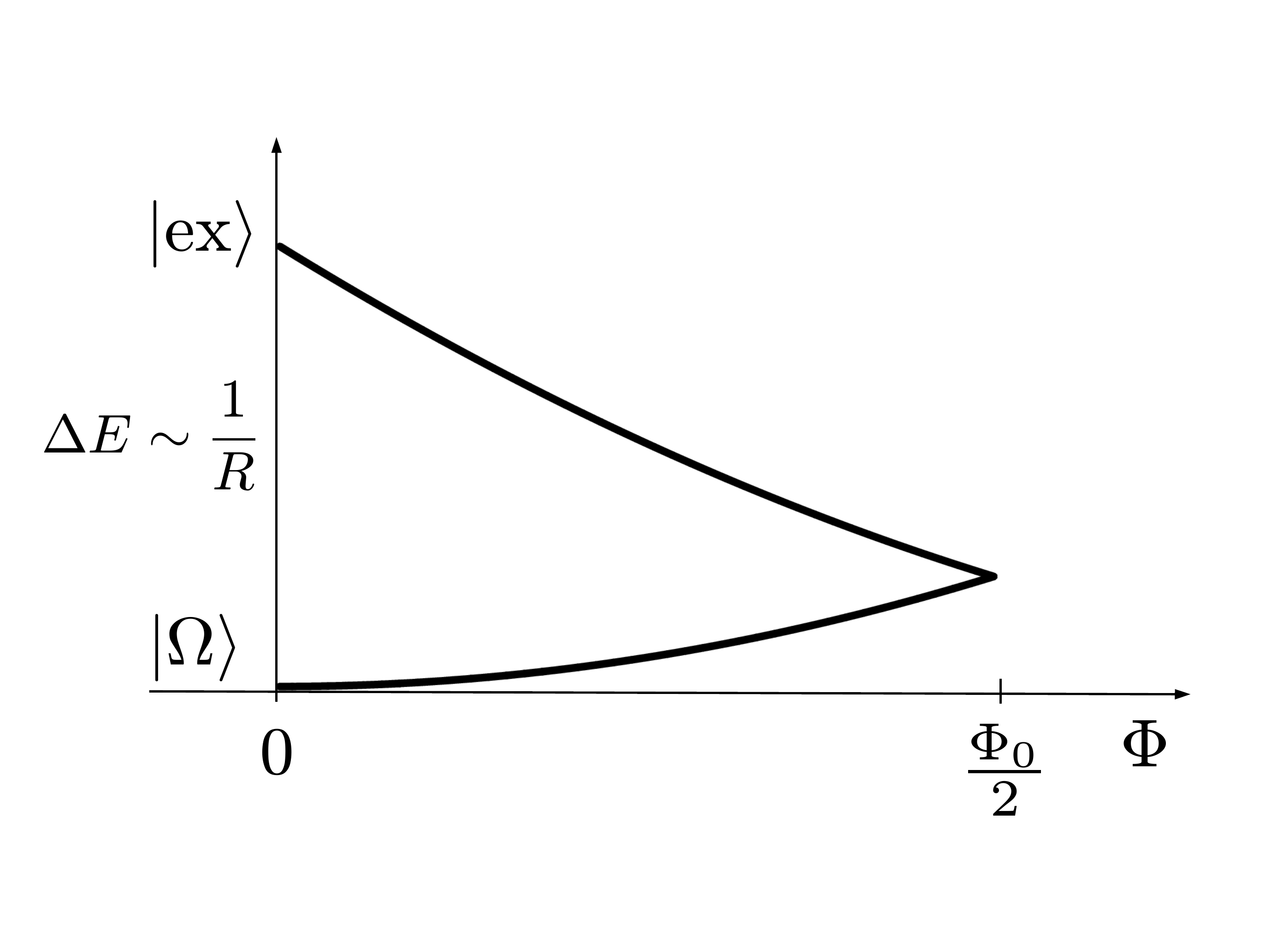}
\caption{Kramers degeneracy at half flux.}
\label{kane-flux}
\end{center}
\end{figure}

The Fu-Kane-Mele argument for stability of time-reversal topological
insulators involves three main steps
(see Fig.\,\ref{kane-flux} for the two-dimensional case) 
\cite{Kane-Z2,LS,cr1}:

i) First, the ground state of the system is adiabatically
deformed by adding half magnetic fluxes, so as to create
a neutral spin one-half excitation at the boundary.

ii) Secondly, the Kramers theorem is invoked, saying that
this excitation is part of a doublet that remains degenerate
in presence of any time-reversal invariant interaction.

iii) Finally, the partner state of the doublet is evolved back to
zero added flux, where it is found to be an excited state 
with energy $O(1/R)$, where $R$ is the size of the system,
thus proving that there is no mass gap in the thermodynamic limit.

In its original implementation for band insulators, this argument
reproduces the $\Z_2$ stability index, because for $N_f$ fermion
species, the created excitation possesses spin $N_f/2$, such that the
Kramers degeneracy is assured for odd $N_f$ only.
Moreover, it extends the validity of the $\Z_2$ index to systems with
interactions and disorder that are time reversal invariant \cite{Kane-Z2}.

In an earlier work on $(2+1)$-dimensional topological insulators
\cite{cr1}, we reformulated the stability argument in terms of
properties of the partition function of edge excitations.  Since the
addition of half flux at the center of the Corbino disk changes the
spatial boundary condition from the (A) to (P), we described the map
between the corresponding partition functions, pertaining to the
Neveu-Schwarz and Ramond sectors, respectively.  We then analyzed the
low-energy states in these partition functions and found the expected
degeneracies and quantum numbers of the excitations mentioned before.

The advantage of this formulation of the stability argument is that it
can be extended to any interacting system for which the partition
function is known.  For $(1+1)$ dimensional edges, this function can
be computed for many models of the quantum spin Hall effect by using
conformal field theory methods \cite{cz}.  Furthermore,
the general structure of the conformal partition function is known
\cite{cv}: this results allowed to extend the $ \Z_2$ stability index
to topological insulators with Abelian and non-Abelian braidings
\cite{cr1}.  In
$(3+1)$ dimensions, we shall first formulate the argument for the
fermionic surface states in this section and then extend it to
bosonic systems in section 4.

\subsubsection{Neveu-Schwarz sector}

The natural boundary conditions for the fermion field are antiperiodic
both in space and time, i.e $(\a_0,\a_1,\a_2)=(1/2,1/2, 1/2)$ in the
partition function \eqref{partiz-funz-3d}.  We call this choice the
Neveu-Schwarz sector, in analogy with $(1+1)$ dimensions.

The low-lying excitations in this sector can be more easily
understood for a rectangular torus, setting
${\boldsymbol{\omega}}_1$ and ${\boldsymbol{\omega}}_2$ along
the Cartesian axes (see Fig.\,\ref{torus}), i.e. $\w_{12}=\w_{21}=0$ 
in (\ref{omegagen}). 
In this case the energy of excitations \eqref{energie} has the form:
\be
\mathcal{E}^{\frac{1}{2} \frac{1}{2}}_{n_1 n_2}= \w_{00}\sqrt{
  \left(n_1 + \frac{1}{2}\right)^2\frac{1}{\w^2_{11}}+\left(n_2 +
  \frac{1}{2}\right)^2\frac{1}{\w_{22}^2}}\ .
\label{NS-spec}
\ee
There are four low-lying degenerate energy levels, for
$(n_1,n_2)=(0,0),(0,-1),$ $(-1,0),$ $(-1,-1)$.
The expansion of the partition function gives:
\begin{align}
\label{exp-AAA}
Z^{F}_{\frac{1}{2},\frac{1}{2} \frac{1}{2}} \sim 
1+\!\!\!\!\!
\sum_{(n_1,n_2)=(0,0),(0,-1),(-1,0),(-1,-1)} \!\!\!\!\!\!\!\!\!\!\!\!\!\!\!\!\!
e^{-2\pi \mathcal{E}_{\vec n} + 2\pi i \mathcal{P}_{\vec n} -  \w_{00}A_0}+
e^{-2\pi \mathcal{E}_{\vec n} - 2\pi i \mathcal{P}_{\vec n} +  \w_{00}A_0}.
\end{align}
Therefore, the low-lying states are the ground state 
plus four particle and four antiparticle
excitations; their fermion parity can be read from
the definition (\ref{spec-f}). In particular, for the ground state,
\be
\label{NS-QF}
(-1)^F \ket{\Omega}_{NS}=
\ket{\Omega}_{NS},
\qquad \quad (-1)^{2S}= (-1)^F.
\ee
In order to discuss the Fu-Kane-Mele stability argument, we need to distinguish
between surface excitations with integer and half-integer spin, and to
this effect, we shall introduce the `spin parity' index $(-1)^{2S}$,
that is equal to the fermion parity of the $(2+1)$-dimensional theory,
$(-1)^{2S}=(-1)^F$ \cite{cr1,cft}.


\subsubsection{Ramond sector}

\begin{figure}[t]
\begin{center}
\includegraphics[width=15cm]{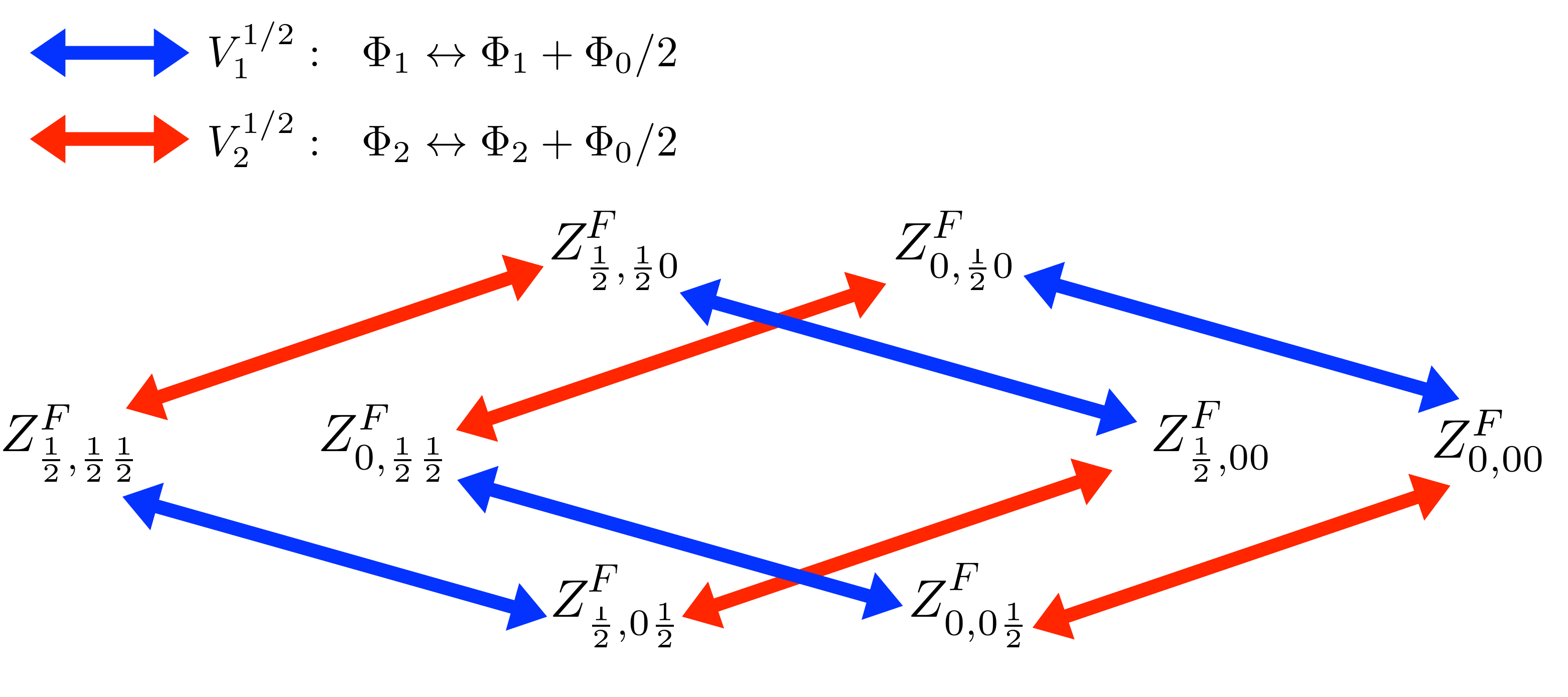}
\caption{Transformations of partition functions 
$Z^F_{\alpha_0,\alpha_1 \alpha_2}$ under addition of half fluxes,
$V^{1/2}_1$ and $V^{1/2}_2$; the indices $\a_{\m}=0$ (resp. $1/2$) 
indicate periodic (resp. antiperiodic)  boundary conditions, 
$\mu=0,1,2$.}
\label{fermifluxes}
\end{center}
\end{figure}

The first step of the stability argument consists on adiabatically
inserting two $\Phi_0/2$ fluxes in the three-dimensional
Corbino geometry (see Fig.\,\ref{T3flussi}). We call
$V^{1/2}_i$, with $i=1,2$, the related transformations. These
insertions modify the quantization of
the momenta $k_x$ and $k_y$, i.e. the spatial boundary conditions $\alpha_i
\rightarrow \alpha_i +1/2$. 
Starting from the Neveu-Schwarz sector
$Z^F_{\frac{1}{2},\frac{1}{2} \frac{1}{2}}$ and applying $V^{1/2}_1$
and $V^{1/2}_2$ we obtain:
\begin{eqnarray}
\label{V-ins}
&&V^{1/2}_1: \Phi_1 \to \Phi_1+ \Phi_0/2, \ \ \ \ \ \ \ \ \ 
Z^F_{\frac{1}{2},\frac{1}{2} \frac{1}{2}}\longrightarrow 
Z^F_{\frac{1}{2},0\frac{1}{2}}, \\
&&V^{1/2}_2: \Phi_2 \to \Phi_2+ \Phi_0/2, \ \ \ \ \ \ \ \ \ 
Z^F_{\frac{1}{2},0 \frac{1}{2}}\longrightarrow Z^F_{\frac{1}{2},00},
\end{eqnarray}
eventually reaching the periodic-periodic sector 
with partition function $Z^{F}_{\frac{1}{2},00}$, 
that will be called the $(2+1)$-dimensional Ramond sector. 
Fig.\,\ref{fermifluxes} shows 
the transformations of all partition functions under half-flux insertions.

In the Ramond sector, the energy and momentum (\ref{Z-E})-(\ref{Z-P})  
are vanishing for $(n_1,n_2)=(0,0)$, i.e
$\mathcal{E}^{0,0}_{0,0} =\mathcal{P}^{0,0}_{0,0}=0$.  Upon expanding the
Ramond partition function, we find four degenerated states, 
\be
\label{Rexp-pf}
Z^F_{\frac{1}{2},00}\propto 1+ e^{-\w_{00}A_0}+e^{\w_{00} A_0}+
e^{-\w_{00}A_0}e^{\w_{00} A_0}+\dots ,
\ee
that we call $\ket{v^{(i)}}$, $i=1,\dots,4$.
Let us analyze their quantum numbers. Two of them have charge $Q=\pm1$,
\be
\label{R-charged}
e^{-\w_{00}A_0} \leftrightarrow \ket{v^{(2)}}, 
\ \ \ \ \ \ \ e^{\w_{00} A_0} \leftrightarrow \ket{v^{(3)}}.
\ee
The other two states are neutral: upon following the evolution of
spectrum while adding the
fluxes, i.e. $\a_i:1/2 \to 0$, we can see that the Neveu-Schwarz
ground state is mapped into the following Ramond state: 
 $\ket{\Omega}_{NS} \to \ket{v^{(1)}}$.
The fourth term is identified as the expected
partner of the Kramers pair, $\ket{v^{(4)}} ={\cal T} \ket{v^{(1)}}$,
\be
\label{R-neutral}
1 \leftrightarrow \ket{v^{(1)}}, 
\qquad
e^{-\w_{00}A_0}e^{\w_{00} A_0} \leftrightarrow \ket{v^{(4)}} , 
\qquad
 Q\ket{v^{(1)}}=Q\ket{v^{(4)}}=0.
\ee

This identification should be checked by evaluating the spin
parity of the four states, that should read:
\begin{align}
&(-1)^{2S}=(-1)^F=-1\qquad {\rm on}\qquad
\ket{v^{(1)}}, \ \ket{v^{(4)}},
\notag\\
\label{R-ind}
&(-1)^{2S}=(-1)^F=1\qquad\ \  {\rm on}\qquad
\ket{v^{(2)}}, \ \ket{v^{(3)}}.
\end{align}
We recall that the fermion number of Ramond states is well understood in
$(1+1)$-dimensions, being the sum of the chiral and anti-chiral
numbers, $F=F_R+F_L$ \cite{cft}.  We now discuss it
in $(2+1)$ dimensions, and show that it is
actually independent of the dimension and the chiral splitting.  The
Fock space identification of these four states in the Ramond sector
follows from the definitions (\ref{vac-cond}) and the previous
arguments:
\be
\label{R-fock}
\ket{v^{(1)}}=\ket{\W}_R, \quad
\ket{v^{(2)}}= a^\dag_{00}\ket{\W}_R, \quad
\ket{v^{(3)}}=b^\dag_{0 0}\ket{\W}_R, \quad
\ket{v^{(4)}}= b^\dag_{0 0}a^\dag_{0 0}\ket{\W}_R, 
\ee
where $\ket{\W}_R$ is the Ramond ground state.
Let us reconsider the normal ordering of the charge and fermion number
given in (\ref{spec-q})-(\ref{spec-f}), starting from the expansion 
around the Fermi level
of a non-relativistic spectrum at finite volume.  In the case of the
Neveu-Schwarz sector, the Fermi level is located in between the empty
and filled states, because the energy spectrum (\ref{NS-spec}) is strictly
positive.  This gives a clear identification of particles and
antiparticles and determines the standard normal-ordering of the relativistic
expressions used in (\ref{spec-q})-(\ref{spec-f}).

In the Ramond sector, there is an ambiguity because some excitations
are exactly located at the Fermi level. 
We shall assume that they are partially filled:
\be
\label{R-fill}
\langle a^\dag_{00}a_{00}\rangle =x,
\qquad 
\langle b^\dag_{00}b_{00}\rangle =1-x,
\qquad\quad 0\le x <1.
\ee
Thus, the normal-ordered expressions of charge and 
fermion number (\ref{spec-q})-(\ref{spec-f}) should be modified in the term 
$(n_1, n_2)=(0,0)$ of the sums, as follows:
\begin{align}
\label{R-Q}
& Q= \sum_{\vec{ n}}  a^{\dag}_{\vec{ n}}a_{\vec{ n}}-
b_{\vec{ n}}^{\dag}b_{\vec{ n}} +1- 2x,
\\
\label{R-F}
& (-1)^F=(-1)^{\sum_{\vec{ n}}  a^{\dag}_{\vec{ n}}a_{\vec{ n}}+
b_{\vec{ n}}^{\dag}b_{\vec{ n}}+1}.
\end{align}
Upon further assuming the particle-hole symmetric filling $x=1/2$,
we obtain the quantum number assignments given before in (\ref{R-charged}),
(\ref{R-ind}).

In conclusion, the Ramond states $\ket{\W}_R$ and
$b^\dag_{00}a^\dag_{00}\ket{\Omega}_R$, have vanishing charge and negative
spin-parity and are identified with the Kramers doublet that we were
looking for.  Since the Ramond sector corresponds to a time-reversal
invariant point for the Hamiltonian, this degeneracy is robust to
any invariant perturbation. To conclude the Fu-Kane-Mele stability
argument we return to zero flux: while the Ramond ground
state $\ket{\Omega}_R$ goes back to the Neveu-Schwarz state
$\ket{\Omega}_{NS}$, its Kramers partner flows
into the following excited state,
\be
\label{NS-ex}
\ket{ex}_{NS}\ \leftrightarrow \ 
e^{- 2\pi \mathcal{E}_{-1-1}^{\frac{1}{2}\frac{1}{2}}+ \w_{00}A_0} 
e^{-2\pi \mathcal{E}_{-1-1}^{\frac{1}{2}\frac{1}{2}}-\w_{00} A_0}.  
\ee 
The energy of this excitation is $\mathcal{O}(1/R)$, where $R$ is the
typical dimension of the system; this proves that the spectrum stays
gapless (in the thermodynamic limit) for any time-reversal invariant 
interaction.

We remark that the neutral $S=1/2$ excitation created by adding
half fluxes is a nonperturbative excitation in the fermionic theory
with respect to the Neveu-Schwarz ground state,
\be
\label{spin-field}
\ket{\W}_R = \s(0) \ket{\W}_{NS}.
\ee
\\[-5pt]
In the $(1+1)$-dimensional theory, $\s(x)$ is known as the `spin field'
and its properties are well understood, e.g. within the fermionic
description of the Ising model \cite{cft};
much less is known in $(2+1)$ dimensions, to our knowledge.

The stability of the surface excitations can be related to a
$\mathbb{Z}_2$ anomaly, as in the case of the lower 
dimensional topological system \cite{cr1, Stern-Z2}.  
Indeed, the Neveu-Schwarz and Ramond ground states are
eigenstates of a time-reversal invariant Hamiltonian, but
possess different spin-parity index, i.e.  
\be 
\label{Z2-anom}
(-1)^{2 S}\ket{\Omega}_{NS}= \ket{\Omega}_{NS},
\qquad\quad (-1)^{2 S}\ket{\Omega}_R=- \ket{\Omega}_R.  
\ee 
The index is conserved by time reversal symmetry, 
but changes between two invariant ground states, 
without any breaking of the symmetry either explicit or spontaneous.
Therefore, we interprete this change as a discrete $\Z_2$ anomaly,
which is equivalent to the $\Z_2$ index of stability.


\subsection{Modular transformations}
\label{mod-transf}

In this section we study the behavior of the eight partition functions
under the discrete changes of coordinates that map the three-torus into 
itself. The pattern of transformations will further characterize the 
different sectors.
Moreover, we shall associate the stability of topological insulators
to the impossibility of writing a modular invariant partition function
that is consistent with all physical requirements.
These results are close analogs of the $(2+1)$-dimensional
ones \cite{cr1}. 

In the following we set $A_0=0$ for  simplicity, and rewrite 
the partition function \eqref{partiz-funz-3d} as follows:
\begin{equation}
Z^F_{\alpha_0,\alpha_1 \alpha_2}= e^{-VF_0}\prod_{n_1,n_2\in
  \mathbb{Z}} \bigg| 1-\exp\left( -2 \pi \mathcal{E}_{n_1,n_2}^{\a_1\a_2}
+2\pi i \mathcal{P}_{n_1,n_2}^{\a_1\a_2} -2 \pi i \a_0 \right)\bigg|^2.
\label{Z-no-A}
\end{equation}

The modular transformations of the torus $\mathbb{T}^3$ express 
the reparameterization of the moduli
$({\boldsymbol{\omega}}_0, {\boldsymbol{\omega}}_1,
{\boldsymbol{\omega}}_2)$, and form the group $SL(3,\mathbb{Z})$. 
Two subgroups $SL(2,\mathbb{Z})$ act on the two-dimensional 
subspaces $(x_0,x_i)$, $i=1,2$, and have generators 
$T_i: {\boldsymbol{\omega}}_0 \to {\boldsymbol{\omega}}_0+
{\boldsymbol{\omega}}_i $ and 
$S_i: {\boldsymbol{\omega}}_0 \to -{\boldsymbol{\omega}}_i$,
${\boldsymbol{\omega}}_i \to {\boldsymbol{\omega}}_0$ \cite{cft}.
The three-dimensional group is generated by $T_1$ and
$U_1=S_1 P_{12}$, where $P_{12}$ is the permutation of spatial vectors, i.e.
$P_{12}: {\boldsymbol{\omega}}_1 \to -{\boldsymbol{\omega}}_2$,
${\boldsymbol{\omega}}_2 \to {\boldsymbol{\omega}}_1$.
The generators $(T_2, S_2)$ are clearly expressed
in terms of $T_1$ and $S_1$ by $T_2=P_{12}T_1 P_{12}$ and $S_2=P_{12}S_1 P_{12}$.

The action of $P_{12}$ on the partition functions
\eqref{Z-no-A} is manifest: $Z^F_{\alpha_0,\frac{1}{2}
  \frac{1}{2}}$ and $Z^F_{\alpha_0,00}$ are left invariant, while the others
exchange in pairs, e.g. $Z^F_{\alpha_0,\frac{1}{2}0} \leftrightarrow
Z^F_{\alpha_0,0 \frac{1}{2}}$.  Therefore it is
sufficient to study the modular transformations given by $T_1$ and
$S_1$ and then apply $P_{12}$ to find the action of the entire
group.

The action of $T_{1}$ is also simply derived from \eqref{Z-no-A}. 
If $\a_1=0$, the partition functions  $Z^F_{\a_0;0 \a_2} $ are invariant.
If $\a_1=1/2$, $T_1$ changes the
temporal boundary conditions from $(A)$ to $(P)$ and viceversa, i.e
$Z^F_{1/2,1/2\, \a_2} \leftrightarrow Z^F_{0,1/2\, \a_2}$.

The action of the transformation $S_1$ requires some
calculations. Following \cite{Ryu-F}, it is useful to choose coordinates
in which the $\boldsymbol{\w}$ matrix is triangular:
\be
\label{o-special}
\boldsymbol{\w}=\begin{pmatrix}
\w_{00} & \w_{01} & \w_{02} \\
0  & \w_{11} & \w_{12} \\
0  &0  & \w_{22}
\end{pmatrix}=
\begin{pmatrix}
2\pi R_0 & -2\pi \a R_1 & -2\pi \g R_2 \\
0  & 2\pi R_1& -2\pi\b R_2 \\
0  & 0 & 2\pi R_2
\end{pmatrix}.
\ee
In this basis, the fermionic partition functions \eqref{Z1} and
\eqref{Z2}, before making the regularization of the vacuum energy,
takes the following form: 
\be
\label{Z-special}
\begin{split}
Z^F_{\alpha_0,\alpha_1 \alpha_2} =&
\prod_{n_2\in\Z}\left\{\prod_{n_1\in\Z}\left| 1-\exp {\cal A}_{\vec n}\right|^2
\exp{\cal B}_{\vec n} \right\}\ ,
\\
{\cal A}_{\vec n}=&
 -2\pi r_{01}\sqrt{[(n_1+\alpha_1)+\beta
    (n_2+\alpha_2)]^2+[r_{12}(n_2+\alpha_2)]^2}
\\ &
+2\pi i \left[ \alpha (n_1+\alpha_1)+(n_2+\alpha_2)(\alpha \beta+\gamma)
 \right]-2\pi i\alpha_0 \ ,
\\ 
{\cal B}_{\vec n} =&  2\pi
  r_{01} \sqrt{\left[ (n_1+\alpha_1)+\beta(n_2+\alpha_2)
      \right]^2+\left[ r_{12} (n_2+\alpha_2) \right]^2}\ ,
\end{split}
\ee
where we split the products on $n_1$ and $n_2$ and introduced 
the two quantities $r_{01}=R_0/R_1$ e $r_{12}=R_1/R_2$. 

The strategy of the calculation \cite{Ryu-F} is to reduce the
partition function to a product of `massive $\Theta$ functions' whose
$S$ transformation is know.  These functions are defined by
\cite{green}:
\ba
\Theta_{[a,b]}(\tau; m) & =& \prod_{n\in \mathbb{Z}} \left| 1- \exp
\left[ -2\pi \text{Im} (\tau) \sqrt{(n+a)^2+m^2}+2\pi i
  \text{Re}(\tau) (n+a)+2\pi i b \right] \right|^2 
\notag\\ 
&& \qquad \times
\exp \left[ 4\pi \text{Im}(\tau) \Delta(m;a) \right] \ ,
\label{massive-TH}
\ea
where $a, b, m \in \mathbb{H}^+$, $\t \in \mathbb{C}$, and  
$\D(m;a)$ is given by,
\be
\label{DELTA-M}
\Delta (m; a)= -\frac{1}{2\pi^2} \sum_{l>0} \int_0^{+\infty} dt ~
e^{-\frac{\pi^2 m^2}{t}-t l^2}\cos (2\pi l a).  \ee 
Identifying:
\ba
\label{tau-def}
&& a=\alpha_1+\beta (n_2+\alpha_2), \quad b=\gamma(n_2+\alpha_2)+\alpha_0,
\quad m=r_{12}(n_2+\alpha_2),
\notag\\
&&\t=-\frac{\w_{01}}{\w_{11}} + i \frac{\w_{00}}{\w_{11}}= \a + i r_{01},
\ea 
the fermionic partition function
\eqref{Z-special} can be rewritten,
\be
Z^F_{\alpha_0,\alpha_1,\alpha_2}= \prod_{n_2 \in \mathbb{Z}}
\Theta_{[\alpha_1+\beta
    (n_2+\alpha_2),\gamma(n_2+\alpha_2)+\alpha_0]}\left(\a + i r_{01};
r_{12}(n_2+\alpha_2)\right).
\label{part-theta}
\ee 

\begin{figure}[t]
\begin{center}
\includegraphics[width=17cm]{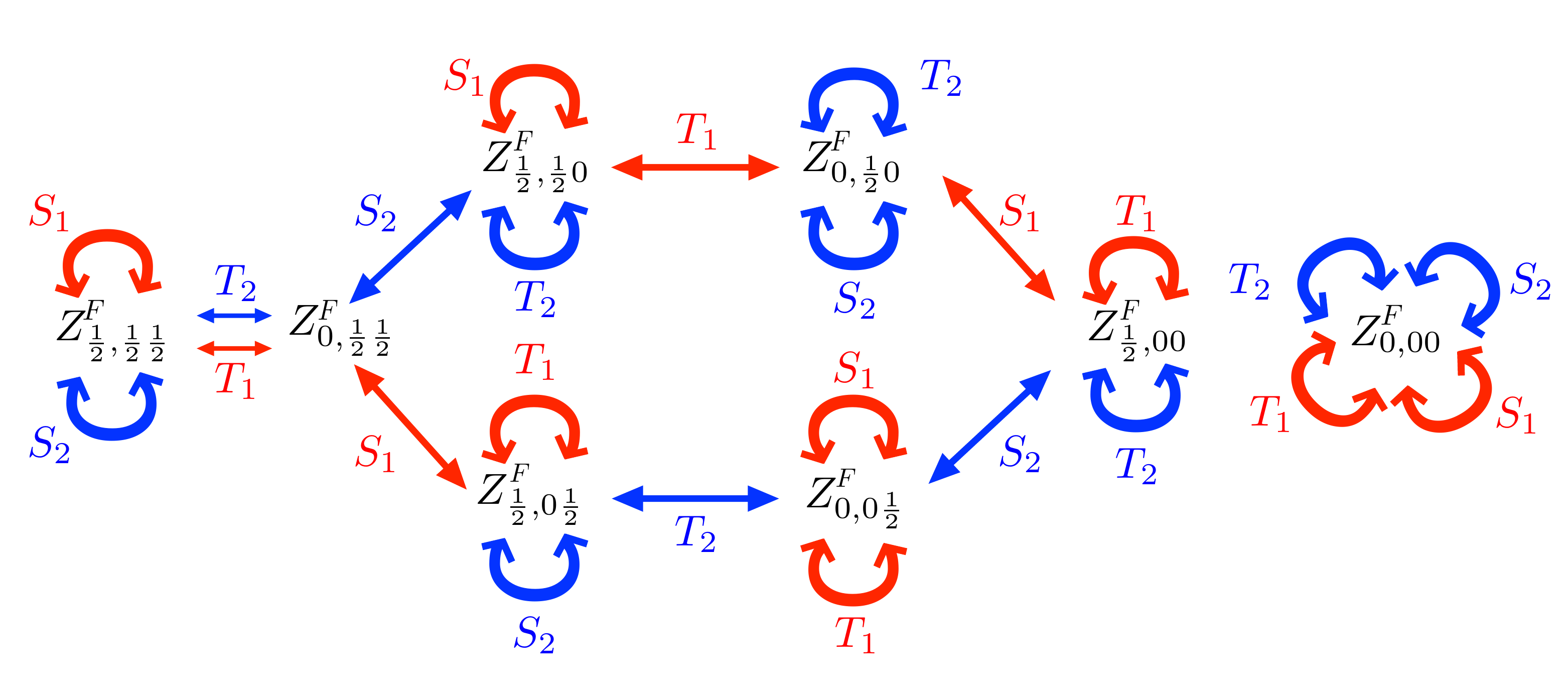}
\caption{Pattern of modular transformations $T_1$, $T_2$, $S_1$,
  $S_2$ for the partition functions $Z^F_{\alpha_0,\alpha_1\alpha_2}$.}
\label{fermi-mod}
\end{center}
\end{figure}

The action of $S_1$  in the basis \eqref{o-special} is:
\begin{align}
\label{Sacts}
&\t \to \ -\frac{1}{\t}, \ \ \ \ \ \ \ \ \a \to
\ -\frac{\a}{\a^2+r^2_{02}}, \ \ \ \ r_{01} \ \to \ \frac{r_{01}}{\a^2
  + r^2_{01}},
\notag \\ &
R_0 \ \to \ \frac{R_0}{|\t|},\ \ \ \ \ R_1
\ \to \ R_1 |\t|, \ \ \ \ \ \ \ \ R_2 \to R_2,\ \ \ \ \ \g \ \to
\ -\b,\ \ \ \ \ \b \ \to \ \g.
\end{align}
We now use the  transformation of the massive $\Theta$ 
function \cite{green},
\be
\label{S-massive}
\Theta_{[a,b]}(\tau;m)= \Theta_{[b,-a]}\left( -\frac{1}{\tau}; m|\tau| \right).
\ee
for each factor in the partition function \eqref{part-theta}, to obtain:
\be
\label{S1-trans0}
\begin{split}
S_1\ :\ & \Theta_{[\alpha_1+\beta
    (n_2+\alpha_2),\gamma(n_2+\alpha_2)+\alpha_0]}(\tau;
r_{12}(n_2+\alpha_2)) \longrightarrow 
\\ &\qquad \Theta_{[\alpha_1+\gamma
    (n_2+\alpha_2),-\beta(n_2+\alpha_2)+\alpha_0]}\left(-\frac{1}{\tau};
r_{12}(n_2+\alpha_2)|\tau|\right) =
\\ &\qquad
\Theta_{[-\alpha_0+\beta(n_2+\alpha_2),\gamma(n_2+\alpha_2) +\alpha_1]}(\tau;
r_{12}(n_2+\alpha_2)),
\end{split}
\ee 
Finally, the $S_1$ transformation of the partition function sis found to be:
\be 
S_1\ :\ Z^F_{\alpha_0,\alpha_1   \alpha_2}
(\boldsymbol{\w}_0,\boldsymbol{\w}_1,\boldsymbol{\w}_2)
\ \to \ Z^F_{\alpha_0,\alpha_1 \alpha_2}
(-\boldsymbol{\w}_1, \boldsymbol{\w}_0, \boldsymbol{\w}_2)=
Z^F_{\alpha_1,\alpha_0 \alpha_2}
(\boldsymbol{\w}_0, \boldsymbol{\w}_1, \boldsymbol{\w}_2).
\label{S1-trans}
\ee
This behavior agrees with the expectations. 
All together, the pattern of modular transformations of the eight 
partition functions $Z^F_{\alpha_0,\alpha_1\alpha_2}$
is shown in Fig.\,\ref{fermi-mod}.


\subsubsection{Stability and modular invariance}
\label{stability-modinv-fermi}

The sum over the eight spin sectors, 
\be
\label{Ztot-F}
Z^{F}_{\text{Ising}}=\sum_{\alpha_0,\alpha_1,\alpha_2=0, \frac{1}{2}}
Z^{F}_{\alpha_0,\alpha_1\alpha_2}, 
\ee 
is found to be invariant under flux insertions $V^{1/2}_1$,
$V^{1/2}_2$ and modular transformations by using the results in
Fig.\,\ref{fermifluxes} and Fig.\,\ref{fermi-mod}.
We can call this modular invariant the `Ising partition
function' being the generalization of a typical partition function of
a statistical model in two dimensions \cite{cr1, cft}.  
However, we have seen that
some sectors, such as the Neveu-Schwarz and Ramond sectors, possess
different values of the ground state spin parity $(-1)^{2S}$ and
cannot be part of the same theory without breaking explicitly the
time reversal symmetry.  Namely, the $\mathbb{Z}_2$ spin parity
anomaly requires that the partition functions stay separate and form a
eight-dimensional vector.  The first component $Z^F_{ \frac{1}{2}
  \frac{1}{2} \frac{1}{2}}$ describes the unperturbed time-reversal
invariant surface system, while the other functions contain excited
states due to changes of electromagnetic and gravitational
backgrounds.

In conclusion, the stability of topological insulators has been
related to the $\mathbb{Z}_2$ spin parity anomaly, which also implies
the modular covariance of the partition function, namely a discrete
gravitational anomaly \cite{cr1, cft}.  We mention that other authors
have been relating the modular covariance of the boundary partition
function to the stability of the topological phase in the bulk
\cite{TI-mod}.

\subsection{Dimensional reduction}

In this section we further characterize the eight fermionic partition
functions by performing a reduction from two to one
spatial dimensions that let us recover well-known expressions
\cite{cft}.

Let us consider the partition functions for a rectangular torus, i.e.
$\w_{12}=\w_{21}=0$, and vanishing scalar potential $A_0=0$, for simplicity.
We perform a dimensional reduction of the Kaluza-Klein type, namely
take the limit $R_2 \to 0$ of the Corbino donut, such that the modes
of energy $O(n_2/R_2)$ are never excited, corresponding to $n_2 \to 0$.
The remaining geometry is that of two-torus in the plane
$(x^0,x^1)$; about the energy spectrum (\ref{Z-E}), there are
two possibilities: i) for periodic boundary condition along $x_2$,
i.e. $\a_2=0$, the spectrum becomes exactly that of the massless
fermion in $(1+1)$ dimensions; ii) for antiperiodic conditions, $\a_2=1/2$,
there remains the constant $1/(4\pi R_2)^2$ that plays the role of a
mass in $(1+1)$-dimensions.

We start from the expression (\ref{Z1})-(\ref{Z2}) before regularization of the
ground state energy, and rewrite it in the coordinates
\eqref{omegagen}:
\ba
Z^F_{\alpha_0,\alpha_1\alpha_2}&=& \prod_{n_2}\left\{
\exp \left[ \frac{2\pi     \w_{00}}{\w_{11}} \sum_{n_1}
  \sqrt{(n_1+\alpha_1)^2+(n_2+\alpha_2)^2\frac{\w_{11}^2}{\w_{22}^2}}
  \right]\right.
\notag\\
&& \quad\times
\prod_{n_1}\bigg| 1-\exp \bigg( -\frac{2\pi \w_{00}}{\w_{11}}
\sqrt{(n_1+\alpha_1)^2+(n_2+\alpha_2)^2\frac{\w_{11}^2}{\w_{22}^2}}
\notag\\ 
&& \quad\quad
 +\frac{2\pi i \w_{01}}{\w_{11}} (n_1+\alpha_1)+
\frac{2\pi i \w_{02}}{\w_{22}}
(n_2+\alpha_2) +2\pi i \alpha_0 \bigg) \bigg|^2 \bigg\}.
\label{2outside1-F}
\ea
The regularized form of the $n_1$ sum in the first exponent is written again
in terms of the  $\D$ function \eqref{DELTA-M}:
\be 
\label{E0-delta}
\sum_{n_1}
\sqrt{(n_1+\alpha_1)^2+(n_2+\alpha_2)^2\frac{\w_{11}^2}{\w_{22}^2}} =
\D\left[\frac{\w_{11}}{\w_{22}}(n_2+\a_2); \a_1\right] \ . 
\ee 
The two-torus is specified by the modular parameter $\t$ in (\ref{tau-def}).
A further convenient simplification is setting $\w_{02}=0$, i.e.
no bending of this torus in three dimensions. 
Altogether, the expression (\ref{2outside1-F}) can be  written again as 
a product over $n_2\in\Z$ of massive $\Th$ functions.
In the limit $\w_{22}\to 0$, the leading behaviour is given by the factor with
$n_2=0$; the dimensional reduction is therefore:
\be
\label{reduced-a2}
Z^F_{\a_0, \a_1 \a_2} \longrightarrow
Z^F_{\a_0, \a_1 | \a_2}=\Theta_{[\a_1, \a_0]}
\left(\t; \ \frac{\w_{11}}{\w_{22}}\a_2 \right).
\ee 
The reduced partition function is denoted by a vertical line 
before the index $\a_2$ of the direction
$x_2\to 0$. We now analyze this expression more explicitly.


\subsubsection{Massless case $\a_2=0$}

In this case, the prefactor $\D(m;a)$ appearing in the theta-function reads:
\be
\label{D-val}
  \D(0;\a_1=0)=-\frac{1}{12}, \ \ \ \ \ \ \D(0; 
\a_1= \frac{1}{2})=\frac{1}{24},
  \ee
 and the ground state energy prefactor becomes,
  \be
\label{F-val}
  \text{exp}\left( 4 \pi \text{Im}\tau \D(0;\a_1)  \right)= 
\begin{cases}
\left( q \bar{q}\right)^{1/12}, \ \ \ \ \ \ \ \a_1=0 ,
  \\ 
  \left(q \bar{q}\right)^{-1/24}, \ \ \ \ \ \a_1=\frac{1}{2},
  \end{cases}
  \ee
  where $q=\exp{(2 \pi i \t)}$.  Remembering that
  $\eta(q)=q^{1/24}\prod_{n=1}^{\infty}(1-q^n)$ is the Dedekind
  function, the reduced partition functions (\ref{reduced-a2}) become
the following expressions:
 \begin{align}
\label{SP-dirac-mass1}
    & Z^F_{\frac{1}{2}, \frac{1}{2} \big | 0} =
\left|\frac{1}{\eta(\t)} \prod_{n=1}^{\infty}(1-q^n) (1+q^{n-1/2})^2 \right|^2
= \left|\frac{1}{\eta(\t)} \sum_{m \in \mathbb{Z}}q^{m^2/2}
\right|^2=Z^{NS}, \\ & \notag \\ 
\label{SP-dirac-mass2}
& Z^F_{0, \frac{1}{2} \big |
  0} = \left|\frac{1}{\eta(\t)} \prod_{n=1}^{\infty}(1-q^n) (1-q^{n-1/2})^2
\right|^2=\left| \frac{1}{\eta(\t)} \sum_{m \in \mathbb{Z}} (-1)^m
q^{m^2/2}\right|^2=Z^{\widetilde{NS}},  \\ & \notag \\ 
\label{SP-dirac-mass3}
& Z^F_{\frac{1}{2}, 0 \big | 0} = \left|\frac{1}{\eta(\t)} 2 q^{1/8}
\prod_{n=1}^{\infty}(1-q^n) (1+q^n)^2 \right|^2=\left|\frac{1}{\eta(\t)}
\sum_{m \in \mathbb{Z}}q^{\left(m+1/2\right)^2/2}\right|^2 =Z^{R},
 \\ &\notag \\ 
\label{SP-dirac-mass4}
& Z^F_{0, 0 \big | 0}= \left|\frac{1}{\eta(\t)}
q^{1/8}\prod_{n=1}^{\infty}(1-q^n) (1-q^n)(1-q^{n-1})
\right|^2=Z^{\widetilde{R}}=0.
\end{align}
These are the well-known partition functions of the
$(1+1)$-dimensional Dirac fermion that describes the edge of the
two-dimensional topological insulator \cite{cr1}.  
In these formulas, we identify the sectors $(AA),(PA),(AP),(PP)$ as 
$NS, \wt{NS},R,\wt{R}$, respectively. We also wrote the bosonic version 
of these expressions \cite{cft}
for later use in Section 4.  The $SL(2,\Z)$ modular transformations of these
partition functions is denoted as the `massless subgroup' shown in
Fig.\ref{fermi-subgroup}.

\subsubsection{Massive case $\a_2=1/2$}

\begin{figure}[t]
\begin{center}
\includegraphics[width=16cm]{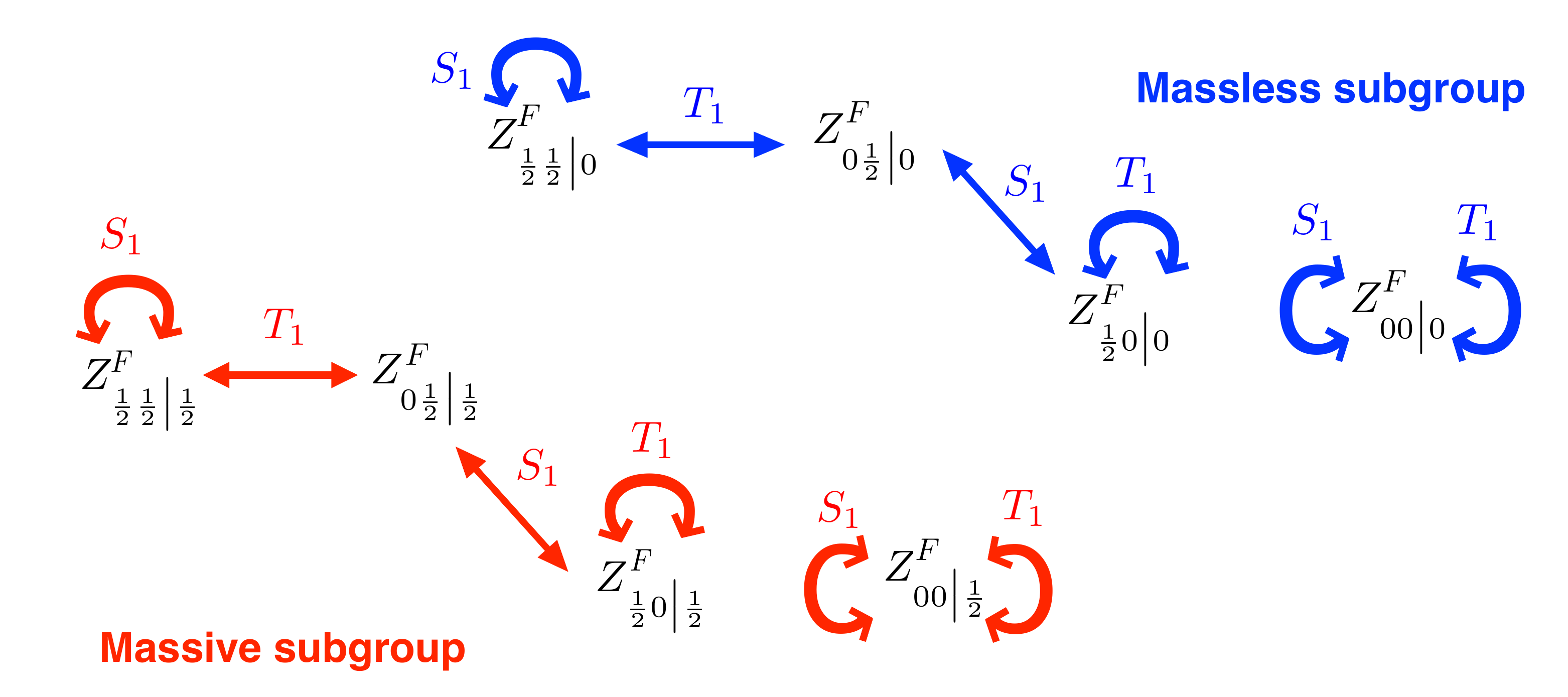}
\caption{Action of the modular group $SL(2,\mathbb{Z})$ over of the
  eight fermionic partition functions
  \eqref{SP-dirac-mass1}-\eqref{SP-dirac-M} dimensionally reduced to the
  $(x^0,x^1)$ plane.}
\label{fermi-subgroup}
\end{center}
\end{figure}

As anticipated, in this case the (large) mass term $M=\w_{11}/2\w_{22}$ remains
in the energy spectrum in $(1+1)$ dimensions. 
Therefore the reduction leads to the following four partition
functions of the massive fermion, that read:
\ba
& &Z^F_{\frac{1}{2}, \frac{1}{2} \big | \frac{1}{2} } =
\Theta_{[\frac{1}{2}, \frac{1}{2}]}(\t; M),
\ \ \ \ \ \ \ \ Z^F_{0, \frac{1}{2} \big | \frac{1}{2} } =
\Theta_{[\frac{1}{2}, 0]}(\t; M),
 \notag \\ 
&&Z^F_{\frac{1}{2}, 0 \big | \frac{1}{2}} = 
\Theta_{[0, \frac{1}{2}]}(\t; M), 
\ \ \ \ \ \ \ \ \ Z^F_{0, 0 \big | \frac{1}{2} } = 
\Theta_{[0, 0]}(\t; M).
\label{SP-dirac-M}
\ea
Their transformations under the subgroup $SL(2,\mathbb{Z})$ are the same
of those of the massless sector, and are indicated as the `massive
subgroup' in Fig.\ref{fermi-subgroup}.

\section{Bosonic topological insulators}

In this section, we discuss the effective field theory description of
free and interacting topological insulators in $(3+1)$ dimensions
given by the topological BF gauge theory and the associated bosonic surface
theory.  We shall recall some known facts, derive the action
at the boundary, its quantization and the partition function on the
three-torus.  We shall then compare these results with those found in
the previous section for the fermionic theory and discuss the insight
they provide on bosonization in $(2+1)$ dimensions.

The motivations for introducing bosonic degrees of
freedom in $(3+1)$ dimensional topological states are the following:
\begin{itemize}
\item 
The great success of bosonic theories in explaining interacting topological
states in $(2+1)$ dimensions, starting from the original work by Wen
on the fractional quantum Hall effect, that provided a complementary
view to wavefunction approaches \cite{Wen-book}. Furthermore, the conformal
field theories describing edge states and braiding relations are usually
formulated in terms of bosonic fields \cite{cft}.
\item 
In particular, the canonical quantization of the compactified free
boson theory in $(1+1)$ dimensions (the so-called chiral Luttinger
liquid) provides an exact description of interacting Hall edge states
with Abelian fractional statistics \cite{cdtz}; 
this approach can be generalized
to topological insulators by taking pairs of theories with opposite
chirality and spin \cite{LS, cr1}.
\item
Such an approach actually corresponds to the BF bulk theory
\cite{TI-qft}, that can be naturally extended to $(3+1)$ dimensions
\cite{Hansson-BF}, where it accounts for particle-vortex braiding
relations and other topological effects \cite{frac-3d}.
\item
We can thus study the corresponding bosonic surface theory in $(2+1)$
dimensions and the dynamics it can support \cite{Moore-BF}.  Of
course, an exact map between fermions and bosons cannot be achieved
\cite{Luther}; nonetheless, we shall find some physical properties
that do not depend on interactions and can be described exactly.
\item
Finally, there are topological states in $(3+1)$
dimensions that are formulated in terms of bosonic microscopic
degrees of freedom \cite{spt-bose}.
\end{itemize}

\subsection{Hydrodynamic BF effective action}

We consider low-energy matter fluctuations that are described by the conserved
currents $J^\mu$ for quasiparticles and $J^{\mu\nu}$ for vortices: these
can be expressed in terms of two hydrodynamic gauge fields,
\be
\label{curr-def}
J^\mu=\frac{1}{2\pi}\eps^{\mu\nu\r\s}\de_\nu b_{\r\s},\qquad
J^{\mu\nu}=\frac{1}{2\pi}\eps^{\mu\nu\r\s}\de_\r a_\s,
\ee
that are the two-form $b=1/2\, b_{\m\n}\, dx^{\m} \wedge dx^{\n}$
and the one-form $a=a_\mu dx^\mu$. 
The topological effects  in time-reversal 
invariant topological states at energies
below the bulk gap can be described by the following BF 
action, with first order derivatives and 
gauge symmetries $a \to a + d\l$ and $b \to b + d\xi$ 
\cite{TI-qft,Moore-BF},
\be
\label{BF-abAt}
S_{BF}[a,b, A]= \int_{\mathcal{M}} \frac{K}{2 \pi} b\,
 da +\frac{1}{2 \pi} b\, dA - \frac{\th}{8 \pi^2} \, da
\, dA +a_\mu {\cal J}^\mu +\frac{1}{2}b_{\mu\nu}{\cal J}^{\mu\nu}\ ,
\ee
where $A=A_\mu dx^\mu$ is the electromagnetic background and
$ {\cal J}_\mu$, ${\cal J}_{\mu\nu}$ are sources for quasiparticle and
vortex excitations.
The time reversal transformations of the fields are:
$a_\mu=(a_0,\vec{a})\to (a_0,-\vec{a})$, 
$A_\mu=(A_0,\vec{A})\to (A_0,-\vec{A})$ and 
$b_{\mu\nu}=(b_{0i},b_{ij})\to (-b_{0i},b_{ij})$, for $\mu=(0,i)$.
Thus, the theory is invariant but for the term proportional to $\theta$.
The coupling $K$ is an odd integer for fermionic systems \cite{frac-3d}.

Same features of the BF theory are:
\begin{itemize}
\item For $A_\mu=0$, the solutions of the equations of motion in presence of
the sources, i.e. 
\be
\label{BFeom1}
{\cal J}^{\m\n}=\frac{K}{2\pi}\e^{\m\n\l\r}\de_{\l}a_{\r}\ , \qquad
{\cal J}^{\m}=\frac{K}{4\pi}\e^{\m\l\r\n}\de_{\l}b_{\r\n}\ ,  
\ee 
imply a non-trivial monodromy of quasiparticles around vortices 
in three space dimensions, with Aharonov-Bohm phases
$\varphi=2\pi N_0 N_1/K$, where $N_0,N_1$ are the quasiparticle 
electric charge and the vortex magnetic charge, respectively.
\item 
For ${\cal J}^\mu=0,{\cal J}^{\mu\nu}=0$, one can compute the induced 
action for the electromagnetic background by integrating the
hydrodynamic fields \cite{Moore-BF},
\be
S_{\text{ind}}[A]= \frac{\th}{8 \pi^2K} \int_{ \cal M}
dA\, dA= \frac{\th }{32 \pi^2K} 
\int d^4x ~ \epsilon^{\mu\nu\lambda \rho} F_{\mu\nu}F_{\lambda \rho}\ .
\label{theta1}
\ee 
This the Abelian theta therm already discussed in the previous section 
\eqref{theta}\cite{TI-qft}:
the case $\th=0$ corresponds to the time reversal invariant system,
where bulk and boundary contributions cancel each other (See Section 2.1).
For $\th=\pi$, time reversal symmetry is broken at the surface, 
leading to the induced Chern-Simons term,
\be
\label{CS-ind-K}
S_{\text{ind}}[A]= \frac{ 1}{8 \pi K} \int_{\de \cal M} A\, dA\ ,
\ee
implying a surface quantum Hall effect with filling fraction
$\nu=1/2K$. In particular, for $K=1$ the fermion anomaly 
(\ref{ind-CS}) is recovered. 
\item
This is the first indication that the bosonic theory for $K=1$ matches
the fermionic description, at least for the topological properties. Other
values of $K$ describe interacting theories with quasiparticle-vortex 
braiding statistics.
\item 
For manifolds $\cal M$ with a boundary, an additional surface action
should be introduced to compensate for the gauge non-invariance of 
the BF theory (\ref{BF-abAt}) \cite{Moore-BF}. This is:
\be
\label{BF-gauge}
S_{\rm surf}[\z,a,A]= -\int_{\de \mathcal{M}} \frac{K}{2 \pi} \z da +
\frac{1}{2 \pi} \z dA,
\ee
\end{itemize}
where the one-form gauge field $\z=\z_\mu dx^\mu$ absorbs the gauge
transformation of the $b$ field, namely $b\to b+d\xi$ and 
$\z\to\z+\xi$.


\subsection{Surface bosonic theory}
\label{surface-theory}

In this section we discuss the massless excitations at the
surface and introduce two dynamics for them that are compatible with the
bulk BF theory and time reversal invariance.
Note in passing that the $(2+1)$-dimensional boundary 
may also support massive phases with topological excitations, whose 
induced action precisely cancels the Chern-Simons term from the bulk 
\cite{top-surf}. These cases will not be discussed here.

The action (\ref{BF-gauge}) (putting $A_\mu=0$ momentarily)
involves boundary degrees of freedom can be viewed as (singular) gauge 
configurations reproducing the bulk loop observables, namely $b=d\z$ 
and $a=d\f$, where $\z$ and $\f$ are a vector and a scalar field, respectively.
The topological BF theory implies a  vanishing Hamiltonian, i.e.
the static case; after choosing the gauge $\z_0=a_0=0$, 
the action (\ref{BF-gauge}) becomes,
\be
\label{BF-2D}
S_{\rm surf}= \frac{K}{2\pi} \int d^3x\,  \e^{ij}\de_i \z_j\, \dot{\f}.
\ee
This expression shows that there are two scalar degrees of freedom
at the surface,  $\f$ and $\c$,
that are canonically conjugate, being the longitudinal part of 
$a_i=\de_i\f$, and the transverse part of $\z_i=\e_{ik} \de_{k} \c$,
respectively,
\be
\label{bf-conj} 
S_{\rm surf}=\int d^3x\, \pi\, \dot{\f},
\qquad \p=\frac{K}{2 \p} \e^{ij}\de_i \z_j=-\frac{K}{2 \p} \D \c.
\ee
Since the surface excitations possess a relativistic dynamics, we should 
add a Hamiltonian. We first consider the free scalar Hamiltonian
in $(2+1)$ dimensions, as follows \cite{Moore-BF}:
\be
\label{bf-dyn}
S_{\rm surf}\to \int d^3x\left(  \p \dot{\f} - {\cal H}(\pi,\f) \right)=
\int d^3x\left(  \p \dot{\f} -\frac{1}{2 m} \p^2-\frac{m}{2}(\de_i \f)^2
\right) \ .
\ee
In this equation, we introduced a mass parameter for adjusting 
the mismatch of dimensions between bulk and boundary: indeed, the
bulk gauge fields imply the mass dimensions $[\f]=0$ and $[\pi]=2$, which
are different from those of the three-dimensional scalar theory
$1/2$ and $3/2$, respectively. A dimensionless coupling could also be
introduced for the third therm in the action (\ref{bf-dyn}), that would
determine the Fermi velocity of excitations. This is conventionally fixed
to one.
The equations of motion of the action (\ref{bf-dyn}) are:
\be
\label{bf-eom}
\p =m \dot{\f}, \qquad \dot{\pi}=m \Delta \phi\ ,
\ee 
and the Lagrangian form of the action is clearly
\be
\label{boson-m}
S_{\rm surf}=\frac{m}{2}\int d^3x\ \left(\de_\mu\f\right)^2 .
\ee
The Hamiltonian equations of motion (\ref{bf-eom}) can be recast into a duality
relation between the boundary scalar and vector fields, that can be written
in covariant form (with $\z_0=0$)\cite{Magnoli-BF}:
\be
\label{dual-1}
\frac{K}{2\pi}\,\e^{\m\n\r} \de_{\n}\z_{\r}=m\, \de_{\m} \f.  
\ee 
This is the just the electric-magnetic duality in $(2+1)$ dimensions:
it plays a role in the bosonization of
$(2+1)$-dimensional fermions through the tomographic representation 
\cite{Luther, Moore-BF} and other approaches \cite{Fradkin-BF}.  
In our setting,
this duality is just the first-order Hamiltonian description of the
relativistic wave equation, that is inherited from the first-order
bulk theory. We also stress that the main motivation for introducing the 
Hamiltonian (\ref{bf-dyn}) is its simplicity. On one side, 
we know that the surface
fermion of the previous section cannot be exactly matched to a free
boson (for $K=1$). On another side, any dynamics of bosonic states
that can model interacting fermions is interesting
to investigate at the present stage of understanding of 
$(3+1)$-dimensional topological insulators.

The coupling to the electromagnetic field $A_\mu$ can be used to 
test the correspondence between boson and fermion theories.
The coupling inherited from the bulk theory is shown in 
(\ref{BF-gauge}) and it amounts to the shift $a_\mu \to a_\mu + A_\mu/K$.
This can be implemented in the symplectic form (\ref{BF-2D}) 
(using the gauge condition $\de_iA_i=0$) and
in the Hamiltonian (\ref{bf-dyn}), thus
leading to the following action:
\be
\label{surf-a}
S_{\rm surf}[\z,\f,A]= 
\int d^3x\left[  \p \left( \dot{\f} -\frac{A_0}{ K} \right)
-\frac{1}{2 m} \p^2-\frac{m}{2}\left(\de_i \f -\frac{A_i}{ K}\right)^2
\right] .
\ee
Upon integrating the scalar fields, we obtain the induced action,
\be
\label{S-ind-m}
S_{\rm ind}^B[A]=\frac{m}{2K^2}\int d^3x \ F_{\mu\nu} 
\frac{1}{\Box}F_{\mu\nu}\ .
\ee
This result should be compared for $K=1$ with the fermionic 
induced action computed in Section 2.1: using Eq. (\ref{Ploop3}),
we disregard the anomalous term cancelled by the bulk and obtain
the expression for $m\to 0$:
\be
\label{S-ind-F}
S^F_{\rm ind}[A]\sim
\int d^3x\, F_{\mu\nu} \frac{1}{\Box^{1/2}}
\left(1+O\left(\frac{m^2}{\Box}\right)\right)F_{\mu\nu}\ .
\ee 

We thus find that the bosonic and fermionic induced actions, (\ref{S-ind-m})
and (\ref{S-ind-F}) do not match, to leading quadratic order in $A_\mu$;
one difference is given by the explicit dimensionful parameter 
of the bosonic theory.
This could be scaled out by the field redefinition,
\be
\label{rescaled}
\widetilde{\f}=\sqrt{m}\f, \qquad \widetilde{\z_i}=\frac{\z_i}{\sqrt{m}},
\qquad \widetilde{\pi}=\frac{1}{\sqrt{m}} \p, 
\ee
but it would not change the induced action (\ref{S-ind-m}),
unless a corresponding shift is considered for the electromagnetic 
coupling. This is not justified because it would imply different
bulk and boundary responses for topological insulators.

We now introduce another Hamiltonian for the bosonic surface states that
is also compatible with the symplectic structure (\ref{BF-2D}) and
relativistic invariance.
Let us reconsider the duality relation between vector and scalar fields
(\ref{dual-1}) and modify it  as follows:
\be
\label{dual-2}
\frac{K}{2\pi}\,\e^{\m\n\r} \de_{\n}\z_{\r}=\Box^{1/2}\, \de_{\m} \f,
\ee
by replacing the mass parameters with a Lorentz invariant non-local operator.
This modified duality corresponds to the following Hamiltonian 
equations of motion:
\be
\label{bf-eom-2}
\pi=\frac{K}{2\pi}\eps^{ij}\de_i\z_j= \Box^{1/2}\dot\f,
\qquad \dot\pi= \Box^{1/2}\Delta\f,
\ee
that follow from the action,
\be
\label{nl-dyn}
S'_{\rm surf}=\int d^3x\left(
\p \dot{\f} -\frac{1}{2} \p \frac{1}{\Box^{1/2}}\p - 
\frac{1}{2} \de_{i}\f\, \Box^{1/2}\de_{i}\f \right) \ .
\ee
In Lagrangian formulation, this reads:
\be
\label{nl-lag}
S'_{\rm surf}=-\frac{1}{2}\int\f\, \Box^{3/2}\f=
\frac{1}{2}\int \left(\de_\mu\wt\f\right)^2, 
\qquad \f=\Box^{1/4}\wt\f,
\ee
that is again the free bosonic theory in the rescaled field variable $\wt\f$. 
The coupling to the electromagnetic field implied by the bulk theory is
still given by the Higgs-like substitution,
$\de_\mu\f\to\de_\mu\f+ A_\mu/K$, leading to the action:
\be
\label{nl-lag-A}
S'_{\rm surf}[\f,A]=\frac{1}{2}\int d^3x
\left(\de_\mu\f-\frac{A_\mu}{K}\right) \Box^{1/2}
\left(\de_\mu\f-\frac{A_\mu}{K}\right).
\ee
Upon integration of the scalar field, we obtain the induced action
$S^{B'}_{\rm ind}$ that is equal to the fermionic expression $S^F_{\rm ind}$
(\ref{S-ind-F}) up to a constant. Therefore, the non-local bosonic
theory (\ref{nl-dyn}) has the same response to weak electromagnetic 
backgrounds as the fermionic theory.

We note that a Lagrangian similar to (\ref{S-ind-F}) was also
introduced in the studies of bosonization of Refs.\cite{Luther}
Although legitimate for a massless theory, a non-local
effective action usually means that further massless excitations have
been integrated out. This may indicate that the dynamics (\ref{nl-dyn})
is incomplete.

In conclusion, we have shown that the surface degrees of freedom
of $(3+1)$ dimensional topological insulators amount to
a Hamiltonian conjugate pair of scalar fields. The
simpler quadratic Hamiltonian for them is not able to reproduce the
fermionic electromagnetic response to leading order, while a modified non-local 
dynamics does work. The two theories are identical on-shell, since
both imply the free wave equation for suitably rescaled field
variables, but may differ in the solitonic excitations.

These open issues are left for further investigations; we remark again that
the analysis in the rest of this paper will deal with properties 
that are independent of the specific dynamics. It is nevertheless
important to stress that the topological data in $(3+1)$ dimensions
given by the BF action (\ref{BF-abAt}) 
do not determine a unique dynamics for the surface states, contrary to the
case of $(2+1)$-dimensional topological insulators.

We finally remark that the local action (\ref{nl-lag}) with coupling
to electromagnetic field given by (\ref{nl-lag-A}) is equivalent to the Abelian
Higgs model in $(2+1)$ dimensions in the deep infrared limit of the
spontaneously broken phase. Namely, the scalar field $\f$
is the Goldstone mode of a complex scalar:
\be
\label{gold-mode}
\F=\r \, {\rm e}^{i\f}\ , \qquad\quad \langle\r^2\rangle = m\ ,
\ee
while the mass parameter $m$ fixes the vacuum expectation value, i.e.
the Higgs field is frozen. We conclude that in cases where the electromagnetic
field could be considered dynamic, we could have a 
superconducting phase at the surface of the topological insulator.
On the other hand, the nonlocal dynamics (\ref{nl-lag-A}) would keep the
photon massless as in the fermionic theory.

\subsection{Canonical quantization of the compactified boson in 
$(2+1)$ dimensions}

In this section we consider the canonical quantization of the
compactified boson with action  (\ref{bf-conj}), (\ref{bf-dyn})
and compute its partition
functions on the three-torus.  We shall pay particular attention to
the properties of solitonic modes of the $\f$ and $\z_i$ fields, in
such a way that they consistently reproduce the topological properties
of the bulk BF theory. We shall follow the analysis of Ref.\cite{Ryu-B}
and extend it in some directions that are rather relevant for the final
result. Some background knowledge can be found in the quantization of
the compactified chiral boson in $(1+1)$ dimensions of Ref\cite{cdtz}.
The quantization of the other non-local theory (\ref{nl-lag-A}) is left for
future investigations.

\subsubsection{Bulk topological sectors and boundary observables}

The quantization of the BF theory (\ref{BF-2D}) on the spatial
three-torus ${\cal M}=\mathbb{T}^3\times \mathbb{R}$, leads to the
topological order of $K^3$ `anyon' sectors, for odd integer values of
the coupling $K$.  The proof of this results is very simple
\cite{Fradkin-book}: one considers the integrals of the
gauge fields $b$ and $a$ on the surfaces and cycles of the torus, 
respectively:
\be
\label{bf-cycles}
\p_{ij}=\int_{\S_{ij}} b,\quad i\neq j, \qquad\quad
q_i=\int_{\g_i} a, \qquad\quad i,j,k=1,2,3.
\ee
These integrals define the global variables $(\p_{ij}(t),q_k(t))$; once 
inserted into the BF action, they become
three pairs of canonically conjugate variables.
The canonical quantization yields the following commutation relations:
\be
\label{CR-cycles}
\left[ \p_{ij}(t),q_k(t)\right]=i\frac{2\pi}{K}\eps_{ijk}\ .
\ee
A basis of holonomies on the torus is given by the operators
$U_i=\exp(iq_i)$ and $V_{ij}=\exp(i\p_{ij})$: they form 
three pairs of $K$-dimensional clock and shift matrices, thus leading to 
a $K^3$-dimensional representation, namely to the
topological order $K^3$.
In the following, we find a corresponding symplectic structure 
among the solitonic modes of the boundary fields $\f, \z_i$,
that are defined on the space-time three-torus.

\begin{figure}[t]
\begin{center}
\includegraphics[width=16cm]{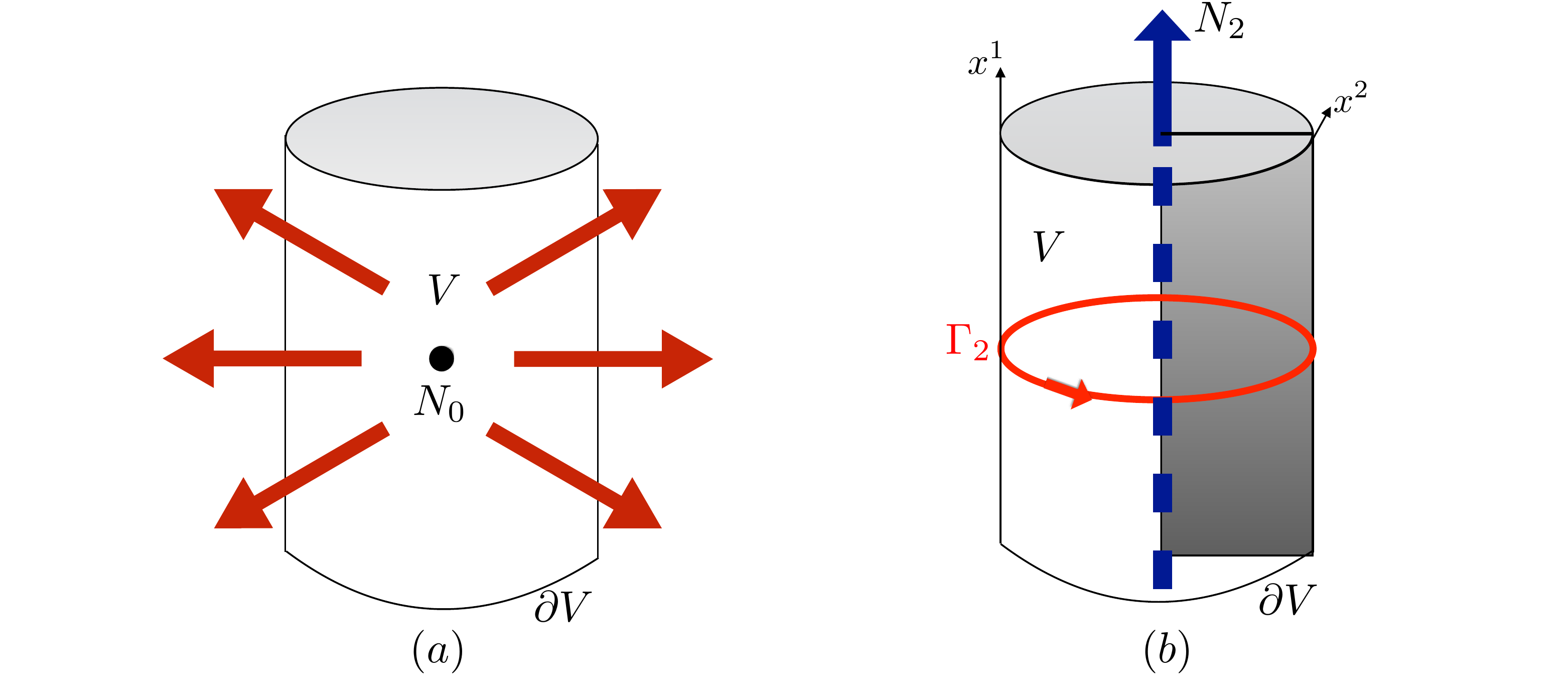}
\caption{(a) Flux due to the electric charge $N_0$ in the
  bulk; (b) Bulk vortex of magnetic charge $N_2$ along $x^1$ and
  closed line $\G_2$ encircling it on the surface.  In grey the branch
  cut surface from the vortex excitation to the boundary.}
\label{bulkflux}
\end{center}
\end{figure}

The relation between bulk and boundary observables can be 
studied on the spatial geometry of the thick two-torus $V=D^2\times S^1$,
whose boundary is the two-torus $\de V =S^1\times S^1$.
We first consider a bulk quasiparticle with charge $N_0$ at rest 
in $\vec{x}=\vec{x}_0$, whose current is
${\cal J}^0(\vec{x})=N_0 \d^{(3)}(\vec{x}-\vec{x}_0)$
(see Fig.\,\ref{bulkflux}(a)). The solution of the
equations of motion (\ref{BFeom1}) leads to a flux of the $b$ field across
a surface enclosing the charge \cite{Ryu-B}, 
that becomes the following expression on the boundary surface:
\be
\label{flux}
\frac{2 \p N_0}{K}=\int_{\de V} d^2x\,\e^{ij}\de_i \z_j.
\ee

A static vortex in the bulk stretched along the non-trivial
cycle with magnetic charge $N_2$ corresponds to the current 
${\cal J}^{01}(\vec{x})=N_2 \d^{(2)}(\vec{x}-\vec{x}_0)$
(see Fig.\,\ref{bulkflux}(b)). 
The equations of motion imply a non-vanishing integral of the $a$ 
field along a closed path encircling the vortex; for the path $\G_2$ 
on the boundary surface, it reads:
\be
\label{winding}
\frac{2 \p}{K} N_2=\oint_{\G_2} dx^2 \de_2 \f \ .
\ee
An analogous relation holds for the other non-trivial cycle of the boundary,
\be
\label{winding-b}
\frac{2 \p}{K} N_1=\oint_{\G_1} dx^1 \de_1 \f\ .
\ee

\subsubsection{Quantization}

The bosonic action (\ref{bf-dyn}) is considered on the space-time geometry
$\de {\cal M}=\de V\times \mathbb{R}= S^1\times S^1 \times \mathbb{R}$.
The conjugate momentum $\p$ is:
\be
\label{pi-2}
\p=\frac{K}{2 \p} \e^{ij}\de_i \z_j\ ,
\ee
and the Hamiltonian equations of motion are,
\be
\label{eom-bf}
\frac{K}{2 \p} \e^{ij}\de_i \z_j=m\dot{\f}\ , \qquad 
\frac{K}{2 \p} \e^{ij}\de_i \dot{\z_j}=m\Delta\f \ .
\ee
The canonical quantization proceeds by expanding the fields in terms of
solutions of the equations of motion, with boundary conditions of the spatial
two-torus specified by the periods $\vec{\w}_i$ 
\eqref{2d-moduli} and dual vectors $\vec{k}_i$ \eqref{k-to-w},
$i=1,2$. 
Let us write the field expansions and then explain them:
\\[-5pt]
\begin{align}
\label{scalar-exp}
& \f(\vec{ x},t) =   \f_0 + 2 \p \L_i \vec{k}_i \cdot \vec{x} +
\frac{K \L_0 t}{mV^{(2)}}
\notag \\ 
&\qquad  +
\frac{1}{ \sqrt{m V^{(2)} } }\sum_{\vec{n} \neq (0,0)} \frac{1}{\sqrt{2
    \En}}\left[a_{\vec{n}}\ \text{e}^{ -i \En t + 2\p i \kn \cdot
    \vec{x}} + a^{\dag}_{\vec{n}}\ \text{e}^{ i \En t - 2\p i \kn
    \cdot \vec{x}}\right], 
\\ & \notag \\ 
\label{vector-exp}
& \z_j(\vec{ x},t) =
\frac{\e_{ji}}{ V^{(2)}}\left( \w_{2i}\g_1 - \w_{1j}\g_2 -\p \L_0x_i\right)
\notag\\ 
&\qquad +\frac{8\pi^2}{K}\sqrt{\frac{m}{V^{(2)}}} \sum_{\vec{n} \neq (0,0)}
\frac{\e_{jm}\left( n_1k_{1m} + n_2 k_{2m}\right)}{\left(2 \En\right)^{3/2}} 
\left[a_{\vec{n}}\ \text{e}^{ -i \En t + 2\p i \kn \cdot
    \vec{x}} + a^{\dag}_{\vec{n}}\ \text{e}^{ i \En t - 2\p i \kn
    \cdot \vec{x}}\right].
\end{align}
These expressions involve oscillating functions specified by 
energies and momenta
\begin{align}
\label{En-oscill}
&\En= 2\p \left| n_1\vec{k}_1 +n_2 \vec{k}_2\right|=\frac{2\pi}{V^{(2)}}	\left|n_1  \vec{\w}_2-n_2 \vec{\w}_1\right|,  \\
\label{kn-oscill1}
&  \vec{k}_{\vec n}=
\left(k_{1\vec{n}}, k_{2\vec{n}}\right)=
\frac{1}{V^{(2)}}\left(n_1\w_{22}-n_2 \w_{12}, \ 
-n_1\w_{21}+n_2 \w_{11}\right)\ .
\end{align}
The field expansions (\ref{scalar-exp}),(\ref{vector-exp}) 
also contain constant and linear terms,
almost unconstrained by the equations of motion, that are needed for
specifying the solitonic modes. Actually, upon inserting these expressions
in the boundary observables (\ref{flux})-(\ref{winding-b}), 
we find the following values
of the $\z_i$ flux and $\de_i\f$ circulations:
\be
\label{flux-quant}
\L_\mu=\frac{N_\mu}{K}, \qquad\quad \a=0,1,2 \ ,
\ee
that explain the normalizations adopted in (\ref{scalar-exp}), 
(\ref{vector-exp}).

The commutation relations between the fields $\f$ and $\pi$,
\be
\label{CR-bf1}
\left[ \f(\vec{x},t),\eps^{ij}\de_i\z_j(\vec{y},t)\right]=
i \frac{2\p}{K}\d^{(2)}(\vec{ x}-\vec{ y}),
\ee
imply the following non-vanishing commutators:
\be
\label{CR-modes1}
    \left[a_{\vec{n}}, a^{\dag}_{\vec{k}}\right]=\d_{\vec{n},\vec{k}} ,
\qquad\quad
 \left[\f_0, \L_0\right]=\frac{i}{K}.
\ee
Moreover, integrating by parts the symplectic term in the action (\ref{BF-2D}),
we can also consider $\z_i$ and $\eps^{ij}\de_j\f$, for $i=1,2$, as two
pairs of coordinates and momenta, leading to 
two further commutation relations:
\be
\label{CR-bf2}
\left[ \z_i(\vec{ x},t), \e_{ij} \de_j \f(\vec{ y},t),\right]=
- i\frac{2 \pi}{k} \d^{(2)}(\vec{ x}-\vec{ y}), \qquad i=1,2.
\ee
These are independent relations for the solitonic modes 
only; they imply the earlier quantizations plus the following ones:
\be
\label{vw-1}
 \left[\g_1, \L_2 \right]=-\frac{i}{K}, \qquad\quad \left[\g_2,
   \L_1\right]=\frac{i}{K}.  
\ee 
These two commutation relations together with that of $\L_0$ in
(\ref{CR-modes1}) represent the bulk degrees of freedom
(\ref{bf-cycles}) within the boundary theory: one quantity in each
pair parameterizes the bulk observable evaluated at the boundary,
i.e. $\L_\mu=(\L_0,\L_1,\L_2)$, while the conjugate variable is a
field zero mode, $(\f_0,\g_2,\g_1)$. After quantization, the
eigenvalues of $\L_\mu$ can be identified with the spectra
(\ref{flux-quant}), that are consistent with the periodicities of the
field zero modes,
\be
\label{compact-r}
\f_0\equiv \f_0+2\pi r, \qquad \g_i\equiv\g_i+2\pi r_i, \quad
i=1,2,
\ee
for compactification radii $r=r_1=r_2=1$. It is also immediate to see
that these periodicities are commensurate with those of the fields
$\f(\vec{x}),\z_i(\vec{x})$ 
winding around the cycles of the torus. In conclusion, 
the $\L_\mu$ spectra  (\ref{flux-quant}) are both suggested by the bulk
theory and consistently obtained by quantization of the boundary
theory. The same result holds in the better-known quantization
of the bosonic theory in $(1+1)$ dimensions at the 
edge of the quantum Hall effect  \cite{cdtz}.
We remark that in both bosonic theories, there are other consistent 
values of the compactification radii, but they would imply 
solitonic spectra that are not related to
the bulk topological data and should be discarded.

The Hamiltonian and the momenta are expressed in terms of
Fock and solitonic operators by substituting the field expansions 
(\ref{scalar-exp})-(\ref{vector-exp}) into standard field expressions of
the bosonic theory (\ref{bf-dyn}). The results are:
\begin{align}
\label{Ham-BF}
&H=\frac{K^2\L^2_0}{2 m V^{(2)}} + \frac{(2 \pi)^2m}{2 V^{(2)}}\left[
  \left(\L_1\w_{22}-\L_2\w_{12} \right)^2 +
  \left(\L_1\w_{21}-\L_2\w_{11} \right)^2 \right]
\notag\\
&\ \ \ +\sum_{\vec{n}\neq
  (0,0)}\En \left( a^{\dag}_{\vec{n}}a_{\vec{n}}+\frac{1}{2}\right),
\\
\label{P1-BF}
&P^1=\frac{2 \p k \L_0}{V^{(2)}}\left(-\L_1\w_{22}+\L_2\w_{12}\right) +
2\p\sum_{\vec{n}\neq
  (0,0)}k_{1\vec{n}}\ a^{\dag}_{\vec{n}}a_{\vec{n}}, \\
\label{P2-BF}
&P^2=\frac{2 \p k \L_0}{V^{(2)}}\left(\L_1\w_{21}-\L_2\w_{11}\right)
+ 2\p\sum_{\vec{n}\neq
  (0,0)}k_{2\vec{n}}\ a^{\dag}_{\vec{n}}a_{\vec{n}},
\end{align}
where the energies $\En$ are momenta $\vec{k}_{\vec n}$ 
are given in \eqref{En-oscill}-\eqref{kn-oscill1}.

\subsection{Bosonic partition functions}

We now compute the partition functions by compactifying the time direction,
i.e. taking the trace over the states.
The sums over the bosonic Fock space and solitonic modes can be done
independently, because their contributions add up in the expressions of
Hamiltonian \eqref{Ham-BF} and  momentum \eqref{P1-BF}, \eqref{P2-BF}.
Thus, the partition function can be factorized into solitonic 
and oscillator parts $Z^{(0)}$ and $Z_{HO}$, respectively
\be
\label{prod-0H0}
Z^B=Z^{(0)}  Z_{HO}.
\ee
The straightforward calculations of the traces on the spectra (\ref{Ham-BF})
-(\ref{P2-BF}) is done by using the coordinates (\ref{2d-moduli}) 
and (\ref{k-to-w}), then
the resulting expressions are written in covariant $(2+1)$-dimensional
notation as functions of the moduli $\boldsymbol{\w}_\mu$, leading to:
\be
\label{Zhocov}
Z_{HO}=\exp (F)
 \prod_{(n_1,n_2)\neq(0,0)} \bigg( 1-\text{exp}\left(-2\p
\mathcal{E}_{\vec{n}}+2 \p i \mathcal{K}_{\vec{n}} \right)
\bigg)^{-1},
\ee 
and  
\begin{align}
 \label{Z0-BFcov}
Z^{(0)}=  \sum_{\L_\mu\in \Z^3/K} \exp \bigg[
& -\frac{V^{(3)}}{|\boldsymbol{\w}_1\times \boldsymbol{\w}_2|^2} 
\left( \frac{K^2\L^2_0}{2 m}
+2 \p^2 m |\L_1 \boldsymbol{\w}_2-\L_2\boldsymbol{\w}_1|^2 \right)
\notag \\ 
& -\frac{i 2 \p K \L_0}{|\boldsymbol{\w}_1\times
  \boldsymbol{\w}_2|^2}\left(\boldsymbol{\w}_1\times
\boldsymbol{\w}_2\right)\cdot \left( \L_1\boldsymbol{\w}_0\times
\boldsymbol{\w}_2 -\L_2 \boldsymbol{\w}_0\times
\boldsymbol{\w}_1\right) \bigg],
 \end{align}
with
\begin{align}
\label{defEPF}
    &  F = \frac{V^{(3)}}{4 \pi} \sum_{(n_1,n_2) \neq(0,0)}
\frac{1}{|n_1\boldsymbol{\w}_2 -n_2\boldsymbol{\w}_1|^3},
\qquad 
\mathcal{E}_{\vec{n}} = V^{(3)}
\ \frac{|n_1\boldsymbol{\w}_2
  -n_2\boldsymbol{\w}_1|}{|\boldsymbol{\w}_1 \times
  \boldsymbol{\w}_2|^2}, 
\\ &
\mathcal{K}_{\vec{n}} =
\frac{(\boldsymbol{\w}_1 \times
  \boldsymbol{\w}_2)}{|\boldsymbol{\w}_1\times \boldsymbol{\w}_2|^2}
\left(n_1\boldsymbol{\w}_0\times \boldsymbol{\w}_2 -n_2
\boldsymbol{\w}_0\times \boldsymbol{\w}_1\right). 
\end{align}
The vacuum energy $F$ is regularized by analytic continuation
as in the fermionic case \cite{cc}. These results
are in agreement with those found in Ref.\cite{Ryu-B}.

%
\subsubsection{Spin sectors of the bosonic theory}
The experience with topological insulators in $(2+1)$
dimensions and bosonization of edge excitations \cite{cr1}
suggests that the partition function just
found (\ref{Zhocov})-(\ref{Z0-BFcov}) should possess the following
properties:
\begin{itemize}
\item 
$Z^B$ should split into the sum over $K^3$ terms, each one pertaining
  to an anyon sectors with given fractional values of the charges of
  the theory.
\item
Further partition functions should be found that correspond to
different quantizations of the solitonic modes, and could be
associated to the eight fermionic spin sectors of the three-torus.
\item
As in the fermionic case, these eight functions should transform one
into another by adding half-flux quanta and changing modular
parameters.
\end{itemize}

Let us gradually derive these results in the $(3+1)$-dimensional theory.
The anyon sectors can be identified by splitting
the summations over the charge lattice $\L_\m\in \Z^3/K$ in $Z^{(0)}$ 
(\ref{Z0-BFcov}) into integer and fractional values, by substituting:
\be
\label{fractional}
\L_{\m}=M_{\m} + \frac{m_{\m}}{K},\qquad
M_{\m} \in \Z, \qquad\ m_{\m}=0, 1, \cdots, K-1, \qquad \m=0,1,2.
\ee
In this way we get the $K^3$ terms, each one involving summations
over integer-spaced charges only, as follows:
\be
\label{Z-anyon}
Z^{(0)}=\sum_{\L_\mu\in \Z^3/K} \cdots =
\sum_{m_0,m_1,m_2=0}^{K-1}\ \sum_{M_0,M_1,M_2\in\Z}\cdots=
\sum_{m_0,m_1,m_2=0}^{K-1} Z^{(0)\, m_0m_1m_2}.
\ee

In Section 2.3, we formulated the flux insertion
argument for the stability of topological insulators 
in terms of fermionic partition functions.
Starting from the Neveu-Schwarz sector, we added half fluxes through
the donut and obtained the other spin sectors of the theory.
In the bosonic theory, adding fluxes clearly modify the values
of the loop observables (\ref{winding}),(\ref{winding-b}):
for example, one flux $\F_0$ adds one unit of magnetic charge to the
corresponding vortex, causing $N_i\to N_i+1$, $i=1,2$.
This is in agreement with the coupling to $A_\mu$ in (\ref{surf-a}).

\begin{figure}[t]
\begin{center}
\includegraphics[width=15cm]{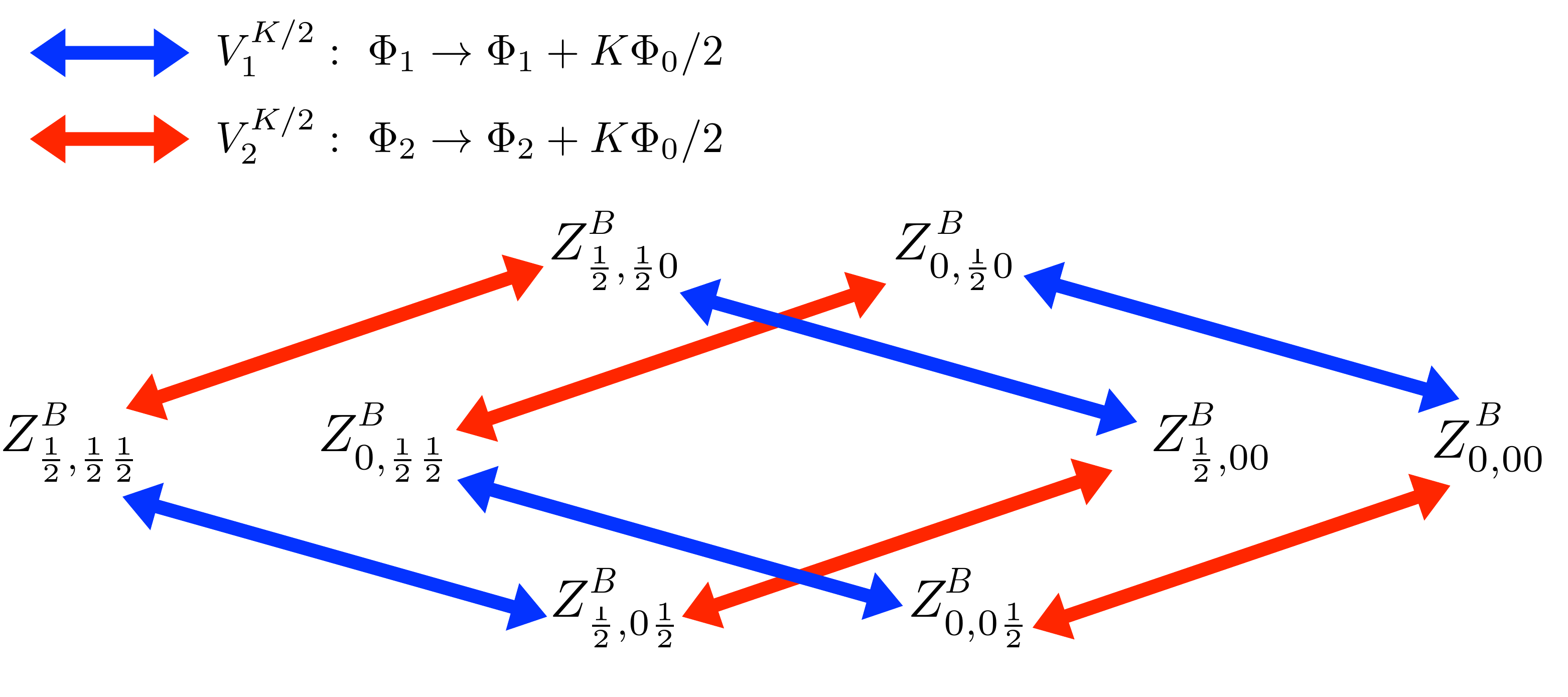}
\caption{$V^{K/2}_1$ and $V^{K/2}_2$ flux transformations on 
the bosonic partition functions $Z^B_{\alpha_0,\alpha_1 \alpha_2}$.}
\label{BF3dflux}
\end{center}
\end{figure}

For $K=1$ adding one flux is clearly a symmetry of the Hamiltonian 
(\ref{Ham-BF}) and of
the partition function \eqref{Z-anyon}, owing to the summation over 
$\L_i\equiv N_i\in\Z$; 
thus, we should consider adding half fluxes by the transformations,
\be
\label{flux-ins}
V^{1/2}_i: \ \ \ \ \Phi_i \ \to \ \Phi_i + \frac{\Phi_0}{2}
\qquad\quad \L_i  \ \to \ \L_i +\frac{1}{2}, \qquad i=1,2.
\ee
that shifts the $\L_i$ summation variables.
For $K>1$ odd, again building on the experience in $(2+1)$ 
dimension \cite{cr1, LS},
we should add a number of fluxes that do not change the anyon sector, i.e. the
fractional value of $\L_i$. Thus, we consider the transformations:
\be
\label{flux-insK}
V^{K/2}_i: \ \ \ \ \L_i=M_i+\frac{m_i}{2}  \ \to \ \L_i +\frac{1}{2}, 
\qquad i=1,2.
\ee
We shall introduce three labels $\a_\mu=0,1/2$, $\mu=0,1,2$ 
to the partition function (\ref{prod-0H0}), as follows:
\be
\label{BF-a}
Z^{B}_{\alpha_0,\alpha_1 \alpha_2}=Z_{HO}\, Z^{(0)}_{\alpha_0,\alpha_1 \alpha_2},
 \qquad\quad \alpha_0,\alpha_1 , \alpha_2=0, \frac{1}{2}.
\ee	
Two of them, $\a_1,\a_2$, specify the half-integer values taken by the
variables $M_1,M_2$ after flux insertions (\ref{flux-insK}), while
$\a_0=1/2$ amounts to adding the sign $(-1)^{K\L_0}$ to the summand in
(\ref{Z0-BFcov}) for reasons that will be clear in the following. Note
that oscillator part $Z_{HO}$ stays invariant.  In conclusion, we have
the following eight partition functions:
\begin{align}
 \label{Z000}
Z^B_{\a_0,\a_1\a_2}=& \sum_{m_\mu\in \Z^3_K} Z_{\a_0,\a_1\a_2}^{B\, m_0 m_1 m_2}
\notag\\
= & Z_{HO} \sum_{m_\mu\in \Z^3_K}  \sum_{M_\mu\in\Z^3}  (-1)^{2\a_0K\L_0}
\notag\\
&\qquad \times\exp \bigg[
 -\frac{V^{(3)}}{|\boldsymbol{\w}_1\times \boldsymbol{\w}_2|^2} 
\left( \frac{K^2\L^2_0}{2 m}
+2 \p^2 m |\L_1 \boldsymbol{\w}_2-\L_2\boldsymbol{\w}_1|^2 \right)
\notag \\ 
&\qquad -\frac{i 2 \p K \L_0}{|\boldsymbol{\w}_1\times
  \boldsymbol{\w}_2|^2}\left(\boldsymbol{\w}_1\times
\boldsymbol{\w}_2\right)\cdot \left( \L_1\boldsymbol{\w}_0\times
\boldsymbol{\w}_2 -\L_2 \boldsymbol{\w}_0\times
\boldsymbol{\w}_1\right) \bigg],
 \end{align}
with parameters,
\begin{align}
\label{Z000-param}
\L_0 =M_0+\frac{m_0}{K},  \qquad & \L_1 =M_1+\frac{m_1}{K}+\a_1, 
\qquad \L_2 =M_2+\frac{m_2}{K}+\a_2,
\notag\\
\a_0,\a_1,\a_2=0,\frac{1}{2}.&
 \end{align}
These partition functions are mapped one into another by the flux 
insertions $V^{K/2}_i$, $i=1,2$ as shown in Fig.\,\ref{BF3dflux}.
A characterization of these functions as the bosonic analogues of
the fermionic spin sectors will become clear in the following
discussion.

\section{Bosonization in $(2+1)$ dimensions}

In this section we focus on the set of eight bosonic partition
functions $Z^{B}_{\alpha_0,\alpha_1 \alpha_2}$ for $K=1$. We show
that they have the same modular transformations and other properties of the
fermionic functions $Z^F_{\alpha_0,\alpha_1 \alpha_2}$. We then argue
that these quantities are actually describing a fermionic theory,
although different from the free theory of Section 2.
Our results provide an exact instance of bosonization
in $(2+1)$ dimensions, namely on the correspondence between (interacting)
fermion and boson theories. It concerns the transformation
properties of the spectrum under changes of backgrounds, that 
are actually independent of the dynamics and thus can be studied 
in the free limit of the two theories.

The fermionic characterization of the bosonic partition functions will be
based on the following properties:
\begin{itemize}
\item
The bosonic $Z^B_{\a_0,\a_1\a_2}$ and fermionic $Z^F_{\a_0,\a_1\a_2}$
transform in the same way under modular transformations and
flux insertions.
\item
Within each sector, they become equal under dimensional reduction to
$(1+1)$ dimensions, where the free bosonic and fermionic theories
match exactly.
\item
The fermion number parity is associated to the states of the bosonic theory, 
and checked under dimensional reduction.
\item
The stability argument for fermionic topological states of Section 2
is reformulated in the bosonic theory for $K=1$ and extended 
to $K>1$, thus proving the stability of fractional topological insulators
with odd integer $K$.
\end{itemize}

\subsection{Modular transformations}

We recall from Section \ref{mod-transf} that the action of the 
modular group $SL(3,\mathbb{Z})$ is described by the generators
$T_1$, $S_1$ and $P_{12}$. Their action on the
oscillator $Z_{HO}$ and solitonic $Z^{(0)}_{\a_0,\a_1,\a_2}$ 
factors of the partition
functions will be described in turn.

Under the $P_{12}$ transformation,
\be P_{12}:
\label{P12}
\ \ \ \ \ \ {\boldsymbol{\omega}}_0 \ \to \ {\boldsymbol{\omega}}_0,
\ \ \ \ \ {\boldsymbol{\omega}}_1 \ \to \ -{\boldsymbol{\omega}}_2,
\ \ \ \ \ {\boldsymbol{\omega}}_2 \ \to \ {\boldsymbol{\omega}}_1, 
\ee
we find that $Z_{HO}$ \eqref{Zhocov} is manifestly invariant. The action
on the solitonic part $Z^{(0)}$ (\ref{Z000}) 
(for $K=1$) is equivalent to the
 relabeling of the variables $(N_1 \to -N_2, \ N_2 \to N_1)$, whose values 
are integer or half integer depending on the values of $\a_1,\a_2$. 
Thus, we find
\be
\label{P12int}
P_{12}: \ \ \ \ Z^{B}_{\a_0,\a_1\a_2} \  \to \ Z^{B}_{\a_0,\a_2\a_1}\ ,
\qquad \a_1,\a_2=0,\frac{1}{2}\ .
\ee
The transformation,
\be 
\label{T1}
T_1: \ \ \ {\boldsymbol{\omega}}_0 \ \to \ {\boldsymbol{\omega}}_0
+ {\boldsymbol{\omega}}_1, 
\ee
leaves again  $Z_{HO}$ invariant. 
The solitonic partitions do not
change under $T_1$ if $M_1$ takes integer values;
for half-integer values, the sums acquire the factor $(-1)^{M_0}$,
thus changing the value of the index $\a_0$ from zero to $1/2$
in $Z^{B}_{\a_0,\a_1\a_2}$ (\ref{Z000}).

The action of the transformation $S_1$ is obtained by following the
same strategy of the fermionic case in Section \ref{mod-transf}
\cite{Ryu-F}.  We first choose the coordinates given in
(\ref{o-special}); the solitonic part of the partition function 
$Z^{(0)}_{0,00}$ in (\ref{Z000}) takes the form ($K=1$):
\begin{align}
\label{Z000-special}
   Z^{(0)}_{0,00} =\sum_{M_\mu\in\Z^3}
\text{exp}&\bigg(-\frac{\t_2 M^2_0}{4 \p m R_2} \ - \ \frac{
     \t_2 m (2\pi)^3}{2 R_2}\left[ R_2^2 \left(M_1+\b M_2 \right)^2 +
     R_1^2M_2^2 \right] \notag \\ &+ 2 \p i \a M_0\left(M_1 +\b M_2
   \right) + 2\p i \g M_0M_2 \bigg),
\end{align}
with parameters defined as,
\be
\label{Z-param}
\t=\t_1+i\t_2=-\frac{\w_{01}}{\w_{11}} +i \frac{\w_{00}}{\w_{11}}=\a
+ir_{01}, \ \ \ \ \ \ \frac{\w_{12}}{\w_{22}}=-\b,
\ \ \ \ \ \ \ \frac{\w_{02}}{\w_{22}}=-\g,
\ee
and $r_{01}=R_0/R_1$. The expressions of the functions for other $\a_\mu$ 
values are found by replacing $M_1 \to M_1 + 1/2$, $M_2 \to M_2 + 1/2$ 
and inserting the factor $(-1)^{M_0}$.
The transformation of this expression is,
\be
 \label{S1-trans2}
S_1: \ \ \ \ \ \ \ Z^{(0)}_{0,00}\left(\t, R_0, R_1, R_2, \b,
\g\right) \ \to \ Z^{(0)}_{0,00}\left(-\frac{1}{\t}, \frac{R_0}{|\t|},
R_1 |\t|, R_2, \g, -\b\right).  
\ee 
The first step is to apply the Poisson resummation formula \cite{cft} on 
the $M_0$ sum, i.e.  
\be
\label{Poisson-form}
\sum_{M_0\in \mathbb{Z}} \text{exp}\left(-\p A M_0^2+2 \p i M_0 B
\right)=\frac{1}{\sqrt{A}}\sum_{M_0^{\prime} \in
  \mathbb{Z}}\text{exp}\left(-\frac{\p}{A}\left( M_0^{\prime}-B
\right)^2 \right).  
\ee
Then, another resummation is done on the $M_1$ index and the result
is recast into the original function 
$Z^{(0)}_{0,00}(\t, R_0, R_1, R_2, \b, \g) $ times the factor $|\t|$.

The oscillator part, in the chosen frame and 
before regularization of the vacuum energy, reads
\begin{align}
\label{ZH0-spec}
Z_{HO}=\prod_{\vec{n} \neq(0,0)}&\bigg[ \left( 1-\text{exp}\left(-2\p
  \t_2 \sqrt{(n_1 +\b n_2)^2 +(n_2 r_{12})^2}-2 \p i (\a n_1 +n_2(\g
  +\t_1 \b)) \right) \right)^{-1} \notag 
\\ & \quad\times
  \text{exp}\left(-\p \t_2 \sqrt{(n_1 +\b n_2)^2 +(n_2 r_{12})^2}
  \right)\bigg].
\end{align}
The product over $(n_1, n_2)$ is separated in the ranges $(n_1 \in\Z
\neq 0, n_2=0)$ and $(n_1 \in \Z, n_2 \in\Z \neq 0)$.  The first
product,  after regularization of the zero point energy,
gives the inverse modulus square of the Dedekind
functions $| \eta(\t)|^{-2}$, where $q=\text{exp}(2 \pi i \t)$.
The second factor can be rewritten in terms of the massive theta
functions introduced in Section 2 (\ref{massive-TH}), 
finally leading to the expression
\be
\label{ZHO-special}
Z_{HO}(\t, R_0, R_1, R_2, \b, \g) =\left| \frac{1}{\eta(q)} \right|^2
\prod_{n_2>0 } \Theta^{-1}_{[\b n_2,n_2
    \g]}(\t,r_{12}n_2).  
\ee
In this form, the $S_1$ transformation, acting on torus parameters
as in (\ref{Sacts}), can be evaluated for each $\Th$ factor. The 
Dedekind function obeys $\eta (-1/\t)=\sqrt{-i \t}\, \eta(\t)$ \cite{cft}, and 
of theta function transforms as given in (\ref{S-massive}). It follows that:
\begin{align}
\label{S1-ZH0}
Z_{HO}\left(-\frac{1}{\t}, \frac{R_0}{|\t|}, R_1 |\t|, R_2, \g,
-\b\right)&=\left| \frac{1}{\eta\left(-1/\t \right)} \right|^2
\prod_{n_2>0 } \Theta_{[\g n_2, -\b
    n_2]}\left(-\frac{1}{\tau}; r_{12}n_2|\tau|\right)\notag
\\ &=\frac{1}{|\t|}\left| \frac{1}{\eta(q)} \right|^2 \prod_{n_2 >0}
   \Theta_{[\b n_2, \g n_2]}(\tau; r_{12}n_2)\notag 
\\&
=\frac{1}{|\t|} Z_{HO}(\t, R_0, R_1, R_2, \b, \g) .
\end{align}

\begin{figure}[t]
\begin{center}
\includegraphics[width=16cm]{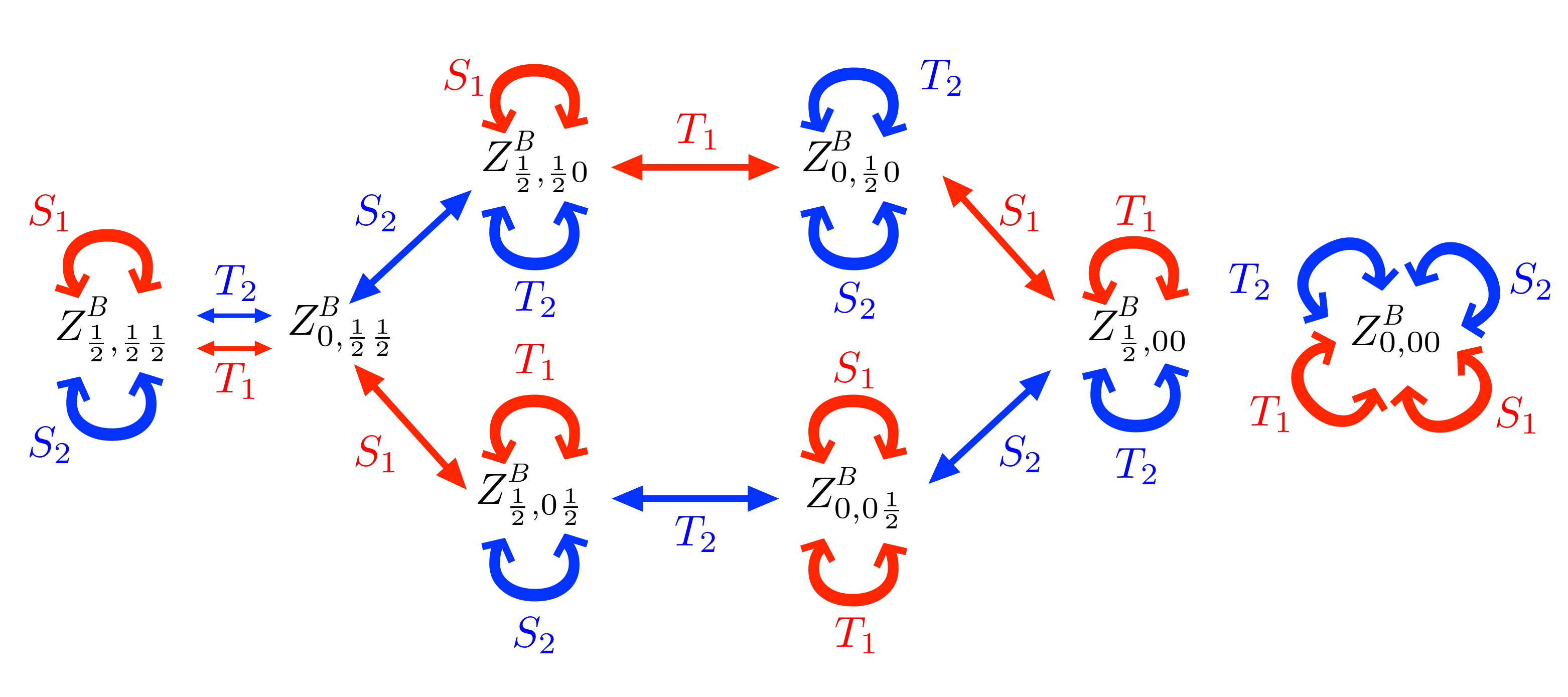}
\caption{Action of the three-dimensional modular group on the bosonic 
partition functions $Z^{B}_{\a_0,\a_1\a_2}$.}
\label{BF3dmod}
\end{center}
\end{figure}

In summary, 
\be
\label{S1-full}
S_1: \ \ \ \ \ Z^{B}_{0, 0 0} \ \to \ Z^{B}_{0,  0 0}.  
\ee
Following the steps illustrated above, we find for the other partition
functions the expected results ($K=1$)
\be 
\label{S1-other}
S_1: \ \ \ \ Z^B_{\alpha_0,\alpha_1
  \alpha_2}(\boldsymbol{\w}_0,\boldsymbol{\w}_1,\boldsymbol{\w}_2)
\ \ \ \longrightarrow \ \ \ Z^B_{\alpha_1,\alpha_0 \alpha_2}
(\boldsymbol{\w}_0, \boldsymbol{\w}_1, \boldsymbol{\w}_2).  
\ee 
Finally, the complete set of modular transformations is shown
in Fig.\,\ref{BF3dmod}. This pattern of transformations 
as well as that for flux insertions in Fig.\,\ref{BF3dflux}
are identical to those of the fermionic theory
 in Figs.\,\ref{fermifluxes} and \ref{fermi-mod}, 
provided that the $Z^B_{\a_0,\a_1\a_2}$ are
put in suitable positions. Before establishing a correspondence term to
term we still need two steps: the dimensional reduction and 
a change of basis that we now discuss.

\subsection{Dimensional reduction}
\label{3to2-bose}

In this section we further characterize the eight bosonic partition
functions $Z^{B}_{\alpha_0; \, \alpha_1 \alpha_2}$ by performing a
dimensional reduction from two to one spatial dimensions, as done in
Section 2.5 for fermions. The mapping to well-know
relations of two-dimensional bosonization \cite{cft} will give useful
informations on the nature of the $(2+1)$-dimensional bosonic sectors.

Let us consider the bosonic partition functions \eqref{Z000} with
$K=1$  for a rectangular torus in the spatial directions, i.e.
$\w_{12}=\w_{21}=0$, for simplicity. We perform again the Kaluza-Klein
dimensional reduction $R_2 \to 0$,
 such that the oscillating and solitonic modes of energy, respectively, 
$O(n_2/R_2)$ and $O(M_2/R_2)$, 
are never excited, corresponding to $n_2, M_2 \to 0$.
Upon setting $\w_{02}=0$, the remaining geometry is that of 
two-torus in the plane $(x^0,x^1)$, with modular parameter $\t$ defined in
(\ref{Z-param}). The reduction of the oscillator part $Z_{HO}$ 
in Eq.\eqref{Z000-special} clearly gives:
\be
\label{ZHO-red}
Z_{HO}(\t)\bigg|_{n_2=0}=\left| \frac{1}{\eta(\t)}  \right|^2 .
\ee

The solitonic factors $Z^{(0)}_{\alpha_0; \, \alpha_1 \alpha_2}$
are similarly expanded for $\w_{22} \to 0$.
Note that the classical action (\ref{bf-dyn}) would vanish in this limit,
as well as the solitonic modes. Thus, we should also vary  the mass 
 that appears as a factor, $m \to \infty$, such that the following product 
remains finite, 
\be
\label{2d-r}
m\w_{22}=r^2/\pi, \qquad {\rm finite}, \qquad \w_{22}\to 0.
\ee
We then find the $(1+1)$-dimensional limit:
\be 
\label{Z000-2d}
Z^{(0)}_{0, 0 |0} = \sum_{M_0, M_1\in
  \Z}q^{\frac{1}{2}\left( \frac{M_0}{2r}+ r M_1\right)^2}
\overline{q}^{\frac{1}{2}\left(  \frac{M_0}{2r}-r M_1\right)^2}.
\ee
This shows that $r$ is the compactification radius of the 
scalar field in $(1+1)$ dimensions, that is fixed to $r=1$ for mapping 
to the free fermion \cite{cft}. Similar expressions are obtained
for the reductions of the other functions $Z^{(0)}_{\a_0,\a_1 |0}$,
by shifting the integers $N_1,N_2$, and adding the sign as indicated in
(\ref{Z000}),(\ref{Z000-param}).

\begin{figure}[t]
\begin{center}
\includegraphics[width=16cm]{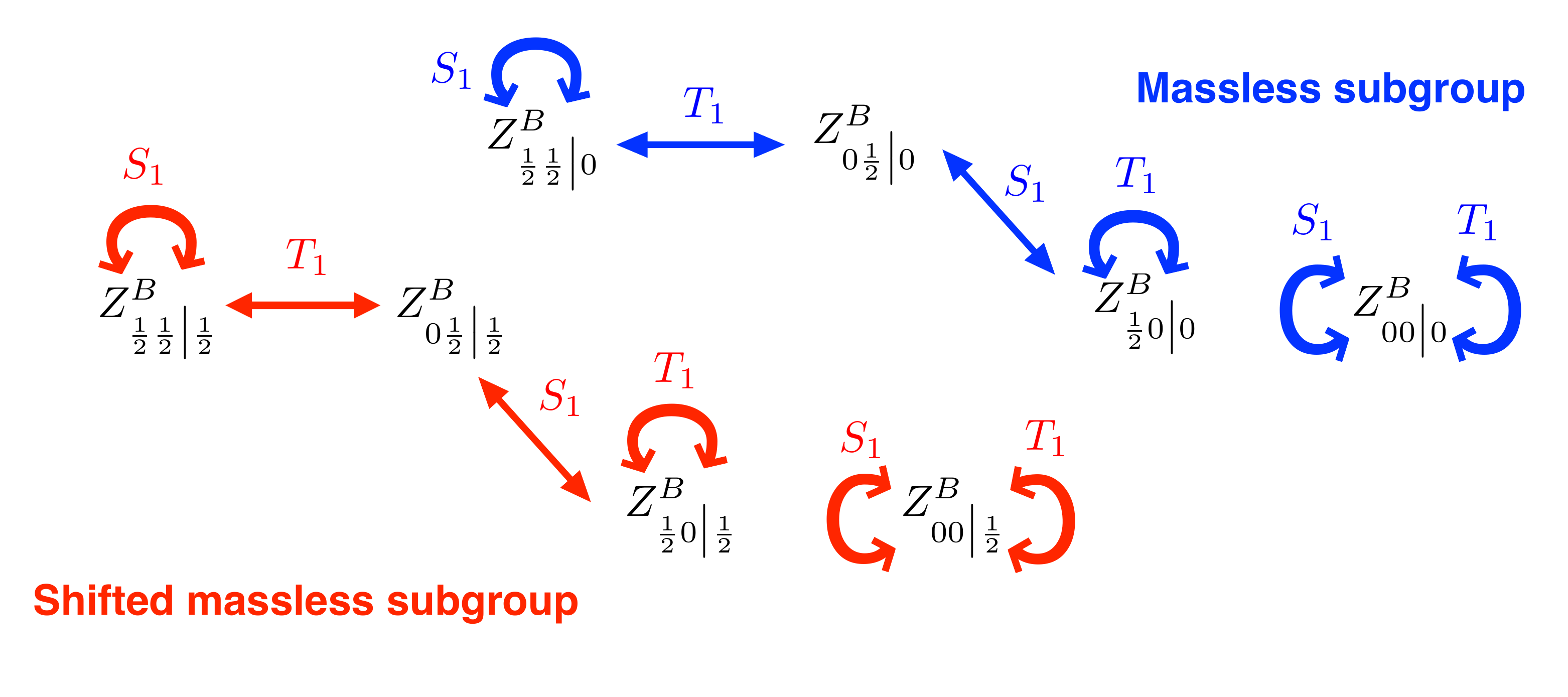}
\caption{Action of the two-dimensional modular group over of the eight
  bosonic partition functions $Z^{B}_{\a_0 \a_1 |\a_2}$ dimensionally
  reduced to the $(x^0,x^1)$ plane.}
\label{bose-subgroup}
\end{center}
\end{figure}

The reduction of the partition functions with $\a_2=1/2$ and $r=1$ leads
to the result
\be
\label{Z-red1}
    Z^{B}_{0, 0 \left|\frac{1}{2}\right.} = 
\left(q \overline{q} \right)^{\w^2_{11}/8\w^2_{22}}
\left| \frac{1}{\eta(\t)}\right|^2
\sum_{M_0, M_1 \in \Z}q^{\frac{1}{2}\left(
  \frac{M_0}{2}+M_1\right)^2}\overline{q}^{\frac{1}{2}\left(
  \frac{M_0}{2}-M_1\right)^2}\ .
\ee
This expression differs from (\ref{Z000-2d}) for the overall ground state
energy $E_0=O(1/R^2_2)\to\infty$, due to the minimal energy of waves
with twisted boundary conditions, that should be subtracted for a
finite limit.  Note that in the fermionic dimensional reduction 
of Section 2.5, the same factor corresponded to a
mass in the dispersion relation (\ref{Z-E}). 
After reduction, the modular transformations of bosonic
functions split again into two $SL(2,\Z)$ 
subgroups, as shown in Fig.\ref{bose-subgroup}.

It is convenient to rewrite the expressions (\ref{Z000-2d}),
(\ref{Z-red1}) in terms
of the fermionic functions for the spin sectors 
$NS,\wt{NS},R,\wt{R}$, by using the bosonization formulae
given in (\ref{SP-dirac-mass1})-(\ref{SP-dirac-mass4}).
The rewriting of (\ref{Z000-2d}) requires the following 
standard manipulations \cite{cft}: splitting in two equal parts, substituting
$(M_0,M_1)=(2\ell, n)$ (resp. $(M_0,M_1)=(2\ell-1, n)$ in the first
(resp. second) part, then replacing 
$\ell + n= \a_0$, $\ell-n=- \overline{\a}_0$ in both of them.
Next, the constraint $\a_0- \overline{\a}_0=2 \ell$ can be
enforced by inserting the projector $\left(1+(-1)^{\a_0+\bar{\a}_0}\right)/2$
into the sums,  finally obtaining:
\begin{align}
\label{Z-red2} 
   Z^{B}_{0, 0 |0} =\frac{1}{2}\left| \frac{1}{\eta}\right|^2
  & \sum_{\a_0, \overline{\a}_0 \ \in \mathbb{Z}}
   \bigg(q^{\frac{1}{2}\a_0^2} \bar{q}^{\frac{1}{2}\overline{\a}_0^2}
   +(-1)^{\a_0+ \overline{\a}_0} q^{\frac{1}{2}\a_0^2}
   \bar{q}^{\frac{1}{2}\overline{\a}_0^2} \notag
   \\ 
&+q^{\frac{1}{2}\left(\a_0 +\frac{1}{2}\right)^2}
   \bar{q}^{\frac{1}{2}\left( \overline{\a}_0 +\frac{1}{2}
     \right)^2}+(-1)^{\a_0+ \overline{\a}_0}q^{\frac{1}{2}\left(\a_0
     +\frac{1}{2}\right)^2} \bar{q}^{\frac{1}{2}\left( \overline{\a}_0
     +\frac{1}{2} \right)^2} \bigg).
\end{align} 
Namely, we find:
\be
\label{Z-BD1}
    Z^{B}_{0, 0|0} =\frac{1}{2}\left( Z^{NS} + Z^{\widetilde{NS}}+
    Z^{R}+ Z^{\widetilde{R}}\right)=Z_{\text{Dirac}}.  
\ee
The other one-dimensional limits of the partition functions become,
following similar steps:
\begin{align}
\label{Z-BD2}
    & Z^{B}_{\frac{1}{2}, 0 \big|0} = \frac{1}{2}\left( Z^{NS} +
Z^{\widetilde{NS}} - Z^{R}-Z^{\widetilde{R}}\right)
\quad\sim Z^{B}_{\frac{1}{2}, 0 \big|\frac{1}{2} } ,
\\
    \label{Z-BD3}
    & Z^{B}_{0, \frac{1}{2} \big|0} = \frac{1}{2}\left( Z^{NS} -
    Z^{\widetilde{NS}} + Z^{R} - Z^{\widetilde{R}}\right)
\quad\sim Z^{B}_{0, \frac{1}{2}  \big|\frac{1}{2} }, 
\\
    \label{Z-BD4}
    & Z^{B}_{\frac{1}{2}, \frac{1}{2} \big|0} = \frac{1}{2}\left( -
    Z^{NS} + Z^{\widetilde{NS}} + Z^{R} - Z^{\widetilde{R}}\right)
\sim Z^{B}_{\frac{1}{2},\frac{1}{2} \big|\frac{1}{2} }.
\end{align}
\\[-5pt] 
In these formulae, we removed the zero-point energies from the 
partition functions with $\a_2=1/2$.


\subsection{Fermion parity in the bosonic theory}

Equations (\ref{Z-BD1})-(\ref{Z-BD4}) show that under dimensional
reduction the eight bosonic partition functions $Z^B_{\a_0,\a_1 \a_2}$
become linear combinations of fermionic sectors $NS,\wt{NS},R,\wt{R}$.
Since $\wt{NS}$ and $\wt{R}$ involve sums of states with
fermion parity $(-1)^{\cal  F}$ in $(1+1)$ dimensions, we find that
the linear combinations obtained by dimensional reduction 
realize the projectors $(1\pm(-1)^{\cal F})$ as in (\ref{Z-red2}).
Namely, they identify $(1+1)$-dimensional
states of given parity, being the same for the Neveu-Schwarz and Ramond
sectors. This result is summarized in the following table:
\be
\label{f-n}
\begin{array}{l|c|c|l|l}
& NS & R & {\rm positive\ Z} &{\rm indefinite\ Z} 
\\
\hline
(-1)^{\cal F} & 1 & 1 & Z^{B}_{0,0|0}\, ,\ Z^{B}_{0,0|\frac{1}{2}} & 
Z^{B}_{\frac{1}{2},0|0}\, , \ Z^{B}_{\frac{1}{2},0|\frac{1}{2} }
\\
(-1)^{\cal F} & -1 & -1 & Z^{B}_{0,\frac{1}{2}|0 }\, ,\ 
Z^{B}_{0,\frac{1}{2}| \frac{1}{2}}
& Z^{B}_{\frac{1}{2},\frac{1}{2}|0 }\, ,\ 
Z^{B}_{\frac{1}{2},\frac{1}{2}| \frac{1}{2}}
\end{array}
\ee
In this table, we also distinguish the 
partition functions that are sums of positive terms,
corresponding to $\a_0=0$, from those with negative signs,
for $\a_0=1/2$, that are due to the factor
$(-1)^{M_0}$ in the definition (\ref{Z000}).

We now discuss the definition of the fermion parity $(-1)^F$
in the bosonic theory in $(2+1)$ dimensions.
Lacking a precise construction of fermionic fields in this theory, we
shall adopt a practical approach. 
A state in $(2+1)$ dimensions will be identified as fermionic if
it matches a $(1+1)$ dimensional fermionic state under dimensional reduction.
We should consider both the limits $x_2\to 0$ given by (\ref{f-n})
and $x_1\to 0$, corresponding to exchanging $\a_1\leftrightarrow\a_2$.
Following this identification, we obtain the table:
\be
\label{f-n-3d}
\begin{array}{ll|c}
& Z & (-1)^{ F}
\\ 
\hline
Z^{B}_{000}\, , &Z^{B}_{\frac{1}{2},00} & 1
\\
Z^{B}_{0,\frac{1}{2}\frac{1}{2}}\, ,& 
Z^{B}_{\frac{1}{2},\frac{1}{2} \frac{1}{2}} &-1
\\
Z^{B}_{0,\frac{1}{2} 0}\, ,& Z^{B}_{0,0 \frac{1}{2} } & /
\\
Z^{B}_{\frac{1}{2}, \frac{1}{2}0}\, ,& Z^{B}_{\frac{1}{2},0 \frac{1}{2} } & /
\end{array}
\ee
The first four partition functions have states with definite
fermion parity, because both dimensional reductions gives the same assignment.
The other four functions have no fermion parity assignment, 
denoted by $(/)$, because the two dimensional reductions give different
results. For example, $Z^{B}_{0,\frac{1}{2} 0} $ would have fermionic 
states according to the dimensional reduction $x_2\to 0$,
i.e. $Z^{B}_{0,\frac{1}{2}| 0} $ in table (\ref{f-n}),
and bosonic states under the reduction $x_1\to 0$, corresponding to 
$Z^{B}_{0,0| \frac{1}{2} } $.
The nature of solitonic states in these four sectors is unclear:
they could correspond to non-local degrees of 
freedom in the fermionic theory that are neither bosonic nor fermionic.

\subsection{Bosonic Neveu-Schwarz and Ramond sectors in $(2+1)$ dimensions}

In the analysis of the fermionic theory, we were able to identify
$(2+1)$-dimensional analogs of the partition functions for 
Neveu-Schwarz and Ramond sectors, that are sums of positive terms and possess
the low-energy expansions:
\be
\label{gs-exp}
NS: \quad Z^F_{\frac{1}{2},\frac{1}{2}\frac{1}{2}}\sim 1+\cdots,
\qquad\quad 
R: \quad Z^F_{\frac{1}{2},00} \sim 2+ \cdots .
\ee
The first state in the $NS$ sector is the ground state and is bosonic,
while the lowest doublet of the $R$ sector is fermionic. These sectors
are mapped one into another by half-flux insertions, 
according to Fig.\,\ref{fermifluxes}; moreover they occupy 
a definite position in the pattern of modular transformations, as
shown in Fig.\,\ref{fermi-mod}.

In the following, we want to identify bosonic functions that possess
these same properties and, moreover, became equal to the
corresponding fermionic functions under reduction to $(1+1)$
dimensions.

According to the patter of bosonic modular transformations shown in
Fig.\,\ref{BF3dmod}, one would be led to the identification
$Z^B_{\a_0,\a_1\a_2}\sim Z^F_{\a_0,\a_1\a_2}$; however, the bosonic
functions go into sums of fermionic sectors under dimensional
reduction, and moreover, the would-be bosonic $NS$ sector
$Z^B_{\frac{1}{2},\frac{1}{2}\frac{1}{2}}$ is not a sum of
positive terms, that is not physically acceptable.  
Note also that dimensional reduction and fermion
parity assignments would favor $Z^B_{000}$ as the candidate $NS$
sector, but this does not have correct modular properties.

The solution to this puzzle is found by considering a change of basis
among the bosonic functions that is an isometry with
respect to the action of the modular group and flux insertions.
Let us first write this transformation and then discuss its features.
The map between the original eight-dimensional basis,
\be
\label{Z-basis}
Z^{B}=(Z^B_{\frac{1}{2}, \frac{1}{2}  \frac{1}{2}}, 
\ Z^B_{0, \frac{1}{2}  \frac{1}{2}},\  Z^B_{\frac{1}{2}, \frac{1}{2} 0},\  
Z^B_{0, \frac{1}{2}  0}, \ Z^B_{\frac{1}{2}, 0 \frac{1}{2}},\  
Z^B_{0, 0 \frac{1}{2}},\  Z^B_{\frac{1}{2}, 00}, \ Z^B_{0, 00}),
\ee
and the new basis $Z^{\prime B}=M Z^{B}$ 
is given by the following matrix:
\be
\label{newbasis}
M=\frac{1}{2}
\begin{pmatrix}
-1 & 1 & 1 & 1 & 1 & 1 & 1 & 1 \\
1 & -1 & 1 & 1 & 1 & 1 & 1 & 1 \\
1 & 1 & -1 & 1 & 1 & 1 & 1 & 1 \\
1 & 1 & 1 & -1 & 1 & 1 & 1 & 1 \\
1 & 1 & 1 & 1 & -1 & 1 & 1 & 1 \\
1 & 1 & 1 & 1 & 1 & -1 & 1 & 1 \\
1 & 1 & 1 & 1 & 1 & 1 & -1 & 1 \\
1 & 1 & 1 & 1 & 1 & 1 & 1 & -1 \\
\end{pmatrix} .
\ee
This transformation leaves invariant the patterns of modular transformations
and flux insertions given in Fig.\,\ref{BF3dmod} and Fig.\,\ref{BF3dflux},
respectively; namely, $M$ commutes with $T_i,S_i, V^{1/2}_i$,
for $i=1,2$. 
Furthermore, it is unique up to exchanges of space coordinates 
$x_1 \leftrightarrow x_2$, that is the up-down reflection of the 
patterns of transformations.

The idea behind the derivation of (\ref{newbasis}) is very simple: the action
of the modular group in Fig.\,\ref{BF3dmod} shows that there are two 
invariants given by the sum of the first seven elements of the multiplet 
 in (\ref{Z-basis}) and by the last element $Z^B_{0,00}$. 
In the new basis, the eighth component should be either the sum or 
the difference of the two invariants; the first choice is not correct because
it would also invariant under flux insertions. The second choice 
yields the other components of the matrix $M$   
by the action of flux and modular transformations.

\subsubsection{Neveu-Schwarz sector}

In the new basis, we are ready to identify
the bosonic analogue of the Neveu-Schwarz sector, that is:
\be
\label{NS-map}
Z^{\prime B}_{\frac{1}{2}, \frac{1}{2} \frac{1}{2}} \ \longleftrightarrow \ 
 Z^F_{\frac{1}{2}, \frac{1}{2} \frac{1}{2}}\ ,
\ee
as suggested by the position occupied in the patterns of 
transformations.
The superposition of bosonic function is given by:
\be
\label{NSboson}
Z^{\prime B}_{\frac{1}{2},  \frac{1}{2}  \frac{1}{2}}=
\frac{1}{2}\left(-Z^{B}_{\frac{1}{2},  \frac{1}{2}  \frac{1}{2}} + 
Z^{B}_{0,  \frac{1}{2}  \frac{1}{2}}+ Z^{B}_{\frac{1}{2},  \frac{1}{2}  0}+
Z^{B}_{0,  \frac{1}{2}  0}+ Z^{B}_{\frac{1}{2},  0 \frac{1}{2}}+
Z^{B}_{0,  0  \frac{1}{2}}+Z^{B}_{\frac{1}{2},  00}+Z^{B}_{0, 00} \right).
\ee
This expression involves sums of terms with positive integer coefficients, 
owing to the presence of the projectors 
$(-(-1)^{M_0}+1)/2$ in the first pair of functions and 
$((-1)^{M_0}+1)/2$ in the other pairs.

The low-energy expansion of the partition functions is done by
inspecting the energy spectrum of solitonic modes \eqref{Ham-BF}
($K=1$),
\be
\label{E-boson}
\mathcal{E}_{M_0 M_1 M_2}^{\a_1 \a_2}=\frac{M_0^2}{2 m V^{(2)}}+
\frac{(2 \pi)^2m}{2 V^{(2)}}\left|
\left(M_1+\a_1\right){\boldsymbol{\omega}}_2
-\left(M_2+\a_2\right){\boldsymbol{\omega}}_1 \right|^2.
\ee
This vanishes for $\a_1=\a_2=0$ and $M_0=M_1=M_2=0$ and
the relative state is found in the term  
$(Z^B_{\frac{1}{2}, 00} + Z^B_{0, 00})/2$ in \eqref{NSboson}.
It gives:
\be
\label{ZB-exp}
 Z^{\prime B}_{\frac{1}{2}, \frac{1}{2} \frac{1}{2}} \sim 1+ \cdots .  
\ee 
This state can be identified with the Neveu-Schwarz,
i.e. unperturbed, ground state of the fermionic system:
it is neutral, since $M_0=0$, and bosonic
owing to the fermion parity assignments (\ref{f-n-3d}) to the
functions $Z^B_{\frac{1}{2}, 00}$ and $ Z^B_{0, 00}$.  The identification
of the ground state is further confirmed by
dimensional reduction. Applying the limits \eqref{Z-BD1}-\eqref{Z-BD4}
to the linear combination in \eqref{NSboson}, we see that the reductions 
for $x_2\to 0$ or $x_1\to 0$ gives the same result by construction,
and reads:
\be 
\label{ZB-NS-red}
Z^{\prime B}_{\frac{1}{2}, \frac{1}{2} \frac{1}{2}} \ \to\ 
\frac{1}{2}\left(Z^{NS}+Z^{\wt NS}+Z^R -Z^{\wt R}\right)
+ \left( \exp (-E_0) \right)Z_{NS}\ ,
\ee
where $E_0$ is the energy shift for $\a=1/2$   in the circle going to zero.
Therefore, the $(2+1)$-dimensional Neveu-Schwarz ground state maps into
its $(1+1)$-dimensional analog, as expected.
In summary, we found the following properties of the bosonic Neveu-Schwarz 
ground state in $(2+1)$ dimensions, 
 \be
\label{B-NS-id}
 1 \leftrightarrow \ket{\Omega}_{NS}, \quad 
H\ket{\Omega}_{NS}= Q\ket{\Omega}_{NS}=0,
\quad (-1)^F\ket{\Omega}_{NS}=(-1)^{2S}\ket{\Omega}_{NS}=\ket{\Omega}_{NS}.
\ee


\subsubsection{Ramond sector}

The $(2+1)$-dimensional Ramond sector is found by doing half-flux insertions
$V_i^{1/2}$, $i=1,2$, that map 
$Z^{\prime B}_{\frac{1}{2},\frac{1}{2}\frac{1}{2}}\ \to \ 
Z^{\prime B}_{\frac{1}{2},00}$.
The corresponding linear combination of bosonic partition functions 
(\ref{Z-basis}) is given by:
\be
\label{Rbose}
Z^{\prime B}_{\frac{1}{2},00}= \frac{1}{2}\left(
Z^{B}_{\frac{1}{2},  \frac{1}{2}  \frac{1}{2}} + 
Z^{B}_{0,  \frac{1}{2}  \frac{1}{2}}+ Z^{B}_{\frac{1}{2},  \frac{1}{2}  0}+ 
Z^{B}_{0,  \frac{1}{2}  0}+ Z^{B}_{\frac{1}{2},  0 \frac{1}{2}}+
Z^{B}_{0,  0  \frac{1}{2}}-Z^{B}_{\frac{1}{2},  00}+Z^{B}_{0, 00} \right).
\ee
This expression is again a sum of terms with  positive integer
coefficients.

Let us analyze the levels with lower energies
 \eqref{E-boson} that are present in
the four pairs $(\pm Z^B_{\frac{1}{2}\a_1\a_2}+ Z^B_{0\a_1\a_2})/2$,
for $\a_1,\a_2=0,1$. Owing to the minus sign in
$(-Z^B_{\frac{1}{2}, 00} + Z^B_{0, 00})/2$, the earlier Neveu-Schwarz
ground state is absent and the lowest states in this pair
occurs for index $M_0=\pm 1$ with energy $O(1/mV^{(2)})$.
The other three pairs of partition functions are inspected for $M_0=0$:
they all possess degenerate level pairs, due to form of the 
energy  (\ref{E-boson}) for $\a_i=1/2$. In particular, let us focus on the
level pair coming from the term
$(Z^B_{\frac{1}{2}, \frac{1}{2}\frac{1}{2}} + Z^B_{0,  \frac{1}{2}\frac{1}{2}})/2$, 
which is,
\be 
\label{Rbose-exp}
Z^{\prime B}_{\frac{1}{2}, 00} \sim \cdots+ \exp\left(
-\mathcal{E}_{000}^{\frac{1}{2}\frac{1}{2}}\right) +
\exp\left(-\mathcal{E}_{0-1-1}^{\frac{1}{2}\frac{1}{2}}\right)+
\cdots,  
\ee 
From the earlier assignments of
fermion parity (\ref{f-n-3d}), the functions $Z^B_{\frac{1}{2},
  \frac{1}{2}\frac{1}{2}}$ and $ Z^B_{0, \frac{1}{2}\frac{1}{2}}$
possess fermionic states, $(-1)^F=(-1)^{2S}=-1$. Thus, the two states
form a Kramers pair under time reversal transformations.
These are not the lowest energy states in the Ramond sector,
but they are fermionic and
one of them, i.e. $\mathcal{E}_{000}^{\frac{1}{2}\frac{1}{2}}$,
is the evolution of the Neveu-Schwarz ground state under the
flux insertions. Thus, these states realize the setting of the Kane-Fu-Mele
stability argument, as shown in Fig.\,\ref{kane-flux}.
The other degenerate pairs belonging to partition functions such as
$Z^B_{0, \frac{1}{2}0}$, do not have a definite fermion parity
assignment; thus, they are not protected by the Kramers theorem.

Summarizing, we have identified in the Ramond sector two degenerate states
$\ket{v}_R, \ket{v'}_R$ with the following properties:
\ba
\exp(-\mathcal{E}_{000}^{\frac{1}{2}\frac{1}{2}})
  \leftrightarrow \ket{v}_{R}, &&
 \text{exp}(-\mathcal{E}_{0-1-1}^{\frac{1}{2}\frac{1}{2}})
\leftrightarrow   \ket{v'}_{R}, \qquad\quad
\mathcal{E}_{000}^{\frac{1}{2}\frac{1}{2}}=
\mathcal{E}_{0-1-1}^{\frac{1}{2}\frac{1}{2}},
\notag \\ 
 Q\ket{v}_{R}=Q\ket{v'}_{R}=0, &&
(-1)^F =-1\quad {\rm on}\quad \ket{v}_{R}, \ket{v'}_{R},
\quad \mathcal{T}\ket{v}_{R}=\ket{v'}_{R}.
\label{R-bose-id}
\ea

The analysis of dimensional reduction in the Ramond sector
\eqref{Rbose} shows that the Kramers pair identified in
(\ref{R-bose-id}) is different from the analog pair relevant in $(1+1)$
dimensions, that is present in the reduction of partition functions
with only one flux insertion, such as $Z^B_{0,\frac{1}{2}0}$.  This
result is consistent with the distinction between strong and weak
three-dimensional topological insulators in three space dimensions
\cite{Kane-Z2}: the surface theories considered here realize the
strong stability case, while the weak topological insulators
correspond to stacks of two-dimensional systems.

\subsection{Stability of bosonic topological insulators}

The previous analysis has shown that the bosonic edge theory 
for $K=1$ possesses
fermionic degrees of freedom: these are not free particles, but
nonetheless their partition functions show the characteristic eight
spin sectors, that are mapped one into the other by the addition of
half fluxes across the two loops of the Corbino geometry and by
modular transformations.

The analysis regarding transformations,
dimensional reductions and fermion parity can be extended to the 
bosonic theory with odd integer values of $K>1$.
A few clarifications are needed:
\begin{itemize}
\item 
The maps between sectors are found by adding $K/2$ fluxes  instead 
of half fluxes, as shown in Fig.\,\ref{BF3dflux}.
\item
Each partition function splits into $K^3$ 
anyon sectors for $m_\mu=0,1,\dots,K-1$ and $\mu=0,1,2$, as shown in
Eq.(\ref{Z000}):
\be
\label{ZB-anyon}
Z^B_{\a_0,\a_1\a_2} = \sum_{m_\mu\in \Z^3_K}
Z^{B\, m_0m_1m_2}_{\a_0,\a_1\a_2}\ ,
\ee
where the indices $m_\mu$ appear as the fractional parts of 
solitonic numbers, $\L_\mu= M_\mu+m_\mu/K$.
\item
The analysis of states and energetics for $K=1$ is also valid for $K>1$, since
it applies to the electron spectrum that is contained in the sub-partition
function (\ref{ZB-anyon}) with $m_\mu=0$ for each spin sector.
\item 
The pattern of modular transformations among the spin sectors is again
given by Fig.\ref{BF3dflux}; within the anyonic sectors, the transformations
are represented by $K^3$-dimensional unitary matrices that will be
described later.
\end{itemize}

The strategy to prove the stability of bosonic (fractional)
topological insulators will be the following: repeat the Fu-Kane-Mele
stability argument for fermionic insulators of Section 2.3, 
addressing the fermionic states identified in the bosonic theory by
the previous analysis.

The ground states $\ket{\W}_{NS}$ of the bosonic Neveu-Schwarz sector 
$Z^{\prime B}_{\frac{1}{2}, \frac{1}{2}\frac{1}{2}}$ in (\ref{ZB-exp}),
(\ref{B-NS-id}), with energy $\mathcal{E}_{000}^{00}$, evolves 
under flux additions $\Delta\Phi_i=K\Phi_0/2$, $i=1,2$
into the Ramond state $\ket{v}_R$, 
with energy  $\mathcal{E}_{000}^{\frac{1}{2}\frac{1}{2}}$
in (\ref{Rbose-exp}), (\ref{R-bose-id}).
Being fermionic, this state possesses a Kramers partner
with energy  $\mathcal{E}_{0-1-1}^{\frac{1}{2}\frac{1}{2}}$;
the pair stays degenerate upon adding 
time-reversal invariant interactions to the Hamiltonian. 
Then, the evolution back to zero flux of the partner state
$\ket{v'}_R$ leads to the following excited state of the Neveu-Schwarz sector,
\be 
\label{NS-ex-B}
\ket{ex}_{NS} \ \leftrightarrow\
\text{exp}(-\mathcal{E}_{0-1-1}^{00}), 
\ee 
whose energy is of order $O(1/R_1,1/R_2)$ (see Fig\,\ref{kane-flux}). 
It follows that the bosonic
spectrum stays gapless (in the thermodynamic limit) in presence of
time-reversal invariant interactions.  This completes the proof of
stability of bosonic topological insulators described by the BF
theory with odd integer coupling $K$.

It is worth stressing the usefulness of the effective field theory
approach  for interacting topological states.
The stability argument originally formulated using band theory
was first translated into the language fermionic surface states and then 
reformulated in terms of properties of 
partition functions \cite{cr1}. Then, the map between fermionic and
bosonic partition functions was used to extend the argument to
interacting topological states within the hydrodynamic approach.

The stability of surface excitations can be again related to a
$\Z_2$ anomaly.  Indeed, the half-flux addition maps states
in the bosonic Neveu-Schwarz and Ramond sector that
possess different spin-parity (fermion parity), according to
the earlier discussion. However, this quantity is conserved by
time-reversal symmetry and no explicit breaking has been introduced.
Therefore, similarly to the fermionic case, we interpret this change
as being a discrete $\Z_2$ anomaly, which is equivalent to the $\Z_2$ index
of stability \cite{cr1}.


\subsubsection{Modular transformations}

We now determine the transformations of the bosonic partition
functions \eqref{Z000} with $K>1$.  The oscillator part of partition
functions $Z_{HO}$ does not depend on $K$ and its transformations were
 described in Section 4.1.  The $K^3$ anyon sectors
$Z_{\a_0,\a_1\a_2}^{B\, m_0 m_1 m_2}$ for each spin sector
carry a unitary linear representation of the modular
group. This is just the generalization of the $K^2$ sectors of
topological insulators in $(2+1)$ dimensions, called 
$K_\l(\t) \ov{K(\t)}_{\l'}$,$\l,\l'\in\Z_K$ in \cite{cz,cr1}; 
the only difference is that there is no chiral factorization.
The action of $T_1$  reads:
\ba
\label{T1-bose}
     T_1: 
Z^{B\, m_0m_1m_2}_{\a_0, 0 \a_2} &\to &
\exp\left( -2 \pi i \frac{m_0m_1}{K} \right)  
Z^{B\,m_0m_1m_2}_{\a_0, 0 \a_2} \ ,
\notag \\
Z^{B\, m_0m_1m_2}_{\frac{1}{2}, \frac{1}{2} \a_2} &\to &
\exp\left( -2 \pi i \frac{m_0m_1}{K} \right)  
Z^{B\,m_0m_1m_2}_{0, \frac{1}{2}\a_2} \ ,
\notag \\
Z^{B\, m_0m_1m_2}_{0,\frac{1}{2} \a_2} &\to &
\exp\left( -2 \pi i \frac{m_0m_1}{K} \right)  
Z^{B\,m_0m_1m_2}_{\frac{1}{2}, \frac{1}{2} \a_2} \ .
\ea
Furthermore, $S_1$ is represented by:
\be
\label{S1-bose}
  S_1: \ \ Z^{B\,m_0m_1m_2}_{\a_0, \a_1 \a_2}  \to  
\sum_{\wt{m}_0, \wt{m}_1\in \Z_K}
\frac{1}{K} \exp\left(2\pi i\frac{\wt{m}_1 m_0+\wt{m}_0m_1}{K}\right)
Z^{B\, \wt{m}_0\wt{m}_1 m_2}_{\a_1, \a_0 \a_2}\ .
\ee
The map between spin sectors for $K>1$ is equal to that of $K=1$
shown in Fig.\ref{BF3dmod}.
As in earlier discussions, the action of $T_2$ and $S_2$ can be found
with the help of the parity $P_{12}$, leading to the matrices:
\begin{align}
\label{T01matrix}
   &  \left(T_2\right)_{m_\mu,\wt{m}_\mu}=
\d^{(3)}_{m_\mu,\wt{m}_\mu}
\exp\left( 2 \pi i \frac{m_0m_2}{K} \right), 
 \notag \\
    & \left(S_2\right)_{m_\mu,\wt{m}_\mu}=
\frac{1}{K} \d_{m_1,\tilde{m}_1} \exp\left(2 \pi i 
\frac{ \tilde{m}_2 m_0 + \tilde{m}_0m_2 }{K}\right),
\end{align}
acting between sectors as in Fig.\,\ref{BF3dmod}.

In Section 3.3 we recalled the quantization of the
global degrees of freedom of the BF theory on the spatial three-torus
${\cal M}=\mathbb{T}^3\times \mathbb{R}$. We then discussed the
relation between bulk and boundary observables, and how the bulk
spectra is reproduced in the quantization of the surface bosonic
theory, through the quantum numbers of solitonic states.  We note that
the matrices $T_1,T_2$ reproduce the statistical phases coming from
braiding anyons around vortex lines \cite{Wen-3loop}.  
This is one instance of the `bulk-boundary' correspondence between 
observables, that has been stressed
in Ref.\cite{Ryu-B} and further investigated for more general hydrodynamic
theories possessing the three-loop braiding statistics 
\cite{Levin-3loop,Ryu-BF2}.

More precisely, in the geometry of the thick spatial two-torus of 
Fig.\,\ref{bulkflux},
the conservation of charge and flux between bulk and boundary implies
that the partition function with anyon indices $(m_0,m_1m_2)$ describes
the edge theory in presence of bulk charge $-m_0$ and bulk
fluxes $(-m_1,-m_2)$.
Modular invariant partition functions are obtained as usual by taking linear 
combinations of anyon sectors. We should consider the case of
vanishing bulk charge $m_0=0$, otherwise there is no symmetry
on exchanging space and time. The following expression 
summing over all fractional values of the fluxes $(m_1,m_2)$,
\be
\label{ZB-modinv}
{\cal Z}^B_{\a_0,\a_1\a_2}=\sum_{m_1,m_2\in \Z_K} 
Z^{\prime B\ 0\,m_1m_2}_{\a_0,\a_1\a_2}\ ,
\ee
is left invariant by the modular transformations, apart from the
usual maps between spin sectors in Fig.\,\ref{BF3dmod}. 
Of course this expression matches
earlier results \cite{cr1} under the dimensional reduction of Section 4.2.

We remark that the stability of the bosonic topological insulators 
is again related to the the impossibility of writing a modular invariant
partition function that is consistent with the physical
requirements. The expression that is invariant under $V^{K/2}_1$,
$V^{K/2}_2$ and the modular group  is
the sum over the eight bosonic spin sectors of (\ref{ZB-modinv}),
\be
\label{ZB-modfull}
{\cal Z}^{B}_\text{INV}=\sum_{\a_0,  \a_1,\a_2=0,\frac{1}{2}}
{\cal Z}^B_{\a_0,\a_1\a_2}.  
\ee 
In analogy with the fermionic case in Section\eqref{mod-transf}, this
partition function is not consistent with time-reversal symmetry
due to the presence of the $\Z_2$ anomaly,  the change of
the spin parity index between the Neveu-Schwarz and Ramond sectors. 
Therefore, time-reversal symmetry requires not to sum over the sectors,
leaving an octet of functions ${\cal Z}^B_{\a_0,\a_1\a_2}$ 
that are modular covariant.

\section{Discussion}

In this paper we have analyzed the fermionic and bosonic theories for
massless surface states of topological insulators in $(3+1)$
dimensions.  We have discussed their effective actions and computed
the partition functions on the three-dimensional torus.  We have found
that their expressions are different in the two theories but transform
in the same way for `large gauge transformations' of the backgrounds,
i.e. for magnetic flux insertions and modular transformations. Of
course, the partition functions become equal under dimensional
reduction, owing to the exact map between free bosons and fermions in
$(1+1)$ dimensions.

In particular:
\begin{itemize}
\item
The theory of the compactified boson with properly quantized 
solitonic spectra displays eight spin sectors on the torus geometry
as in the fermionic theory.
\item
The Neveu-Schwarz (antiperiodic in space) and Ramond (periodic)
sectors of the fermionic theory have been identified in the bosonic
theory, together with their fermion parity assignments.
\item  
The Fu-Kane-Mele argument for stability of topological insulators in
presence of time-reversal invariant interactions has been reformulated
in terms of properties of partition functions for fermionic
surface states.
\item
The stability argument has been extended to bosonic states,
by using the map between fermionic and bosonic partition functions.
\item
The bosonic theory in $(2+1)$ dimensions
corresponds to a yet unknown theory of massless interacting
fermions, and may also contain non-local degrees of freedom.
It describes the surface states of 
topological insulators with fractional Abelian statistics of
odd integer parameter $K$.
\end{itemize}

These results add small bits of information for understanding the
bosonization of relativistic particles in $(2+1)$ dimensions, that are
nonetheless exact properties at the quantum level.

Let us briefly discuss how our study of the bosonic theory is related
with recent conjectures of bosonization in $(2+1)$ dimensions (for $K=1$).
The basic physical picture underlying these correspondences is that of
`attaching flux tubes to particles', that can change the statistics
from fermionic to bosonic and viceversa.
Flux attachment is an experimental fact for non-relativistic
quasiparticles in the fractional quantum Hall effect,
although only understood at the level of ansatz wavefunctions and mean
field theory. One flux per particle can be attached by coupling
matter to a `statistical' gauge field ${\cal A}_\mu$ with Chern-Simons
action and coupling $K=1$: this interaction can be removed by
a (singular) gauge transformation that changes the statistics of
wavefunctions \cite{Fradkin-book}.

Recently, several authors have suggested that flux attachment also
holds for relativistic theories \cite{Son-CF,Seiberg-dual,Tong}.  
For example, a map exists
between the theories of a Dirac fermion and a complex scalar, both
including self-interactions, where the boson is coupled to the
Chern-Simons field for changing statistics \cite{Tong}.
In our analysis of the bosonic theory, we do not include
the Chern-Simons field, but the flux attachment is represented 
by the choice of boundary conditions for the soliton
excitations in the Ramond sector \eqref{flux-ins}-\eqref{flux-insK},
corresponding to half fluxes added along the two spatial cycles of the
torus.
A main difference between our approach and recent bosonization studies
is the compactification of the bosonic field that
is not discussed in \cite{Tong}.  
Let us remark that our attempts to bosonization are physically motivated
by the hydrodynamic theory of topological states of matter,
uniquely fixing the surface degrees of freedom and the coupling to the
backgrounds, as explained in Section 3.  Regarding the dynamics of the
bosonic theory, we pointed out that more than one Hamiltonian
is compatible with the topological data; in Section 3.2 we actually
introduced a second non-local bosonic action \eqref{nl-lag-A}.

Possible developments of our analysis are the following:
\begin{itemize}
\item
The study of the bosonic theory presented in this paper 
can be repeated for the non-local dynamics \eqref{nl-dyn}.
\item
The study of bosonic surface theories can also be extended to models
with two BF hydrodynamic fields that are necessary to describe
the three-loop braiding statistics, following the works 
\cite{Levin-3loop,Ryu-BF2}.
\item
The bosonic theory (\ref{bf-dyn}) discussed in this paper can
be further analyzed for constructing fermionic operators in $(2+1)$
dimensions.  So far, attempts to finding generalized vertex operators
by the so-called tomographic representation have been done at the
semiclassical level \cite{Luther}. A first question in this direction is 
about the realization of the conformal symmetry in the bosonic theories
(\ref{bf-dyn}) and (\ref{nl-dyn}).
\end{itemize}

\bigskip

{\bf Acknowledgments} 

The authors would like to thank A.G. Abanov, T.H. Hansson, C.L. Kane,
G. Palumbo, S. Ryu, K. Schoutens, D. Seminara, D.T. Son, P. Wiegmann
and G.R. Zemba for very useful scientific exchanges. A.C. acknowledge
the hospitality and support by the Simons Center for Geometry and
Physics, Stony Brook, and the G. Galilei Institute for Theoretical
Physics, Arcetri, where part of this work was done.  E.R. acknowledge
the hospitality and support by the Institute of Physics of the
University of Amsterdam and the Delta Institute for Theoretical
Physics, The Netherlands.  The support of the European IRSES grant,
‘Quantum Integrability, Conformal Field Theory and Topological Quantum
Computation’ (QICFT) is also acknowledged.


\end{document}